\documentclass[11pt]{article}

\usepackage[margin=1in]{geometry}
\usepackage[authoryear,round]{natbib}
\usepackage{amsmath,amssymb}
\usepackage{booktabs}
\usepackage{array}
\usepackage{graphicx}
\usepackage{xcolor}
\usepackage{hyperref}
\IfFileExists{placeins.sty}{\usepackage{placeins}}{\newcommand{\FloatBarrier}{}}


\graphicspath{{figures/}}

\hypersetup{
  colorlinks=true,
  linkcolor=blue,
  citecolor=blue,
  urlcolor=blue
}

\newcommand{\modelname}{National Oceanic and Atmospheric Administration/Earth System Research Laboratory (NOAA/ESRL) Flow-following finite-volume Icosahedral Model, version r1p1}
\newcommand{\modelshort}{FIMr1p1}
\newcommand{\methodname}{\modelshort{}-FlowMatch}
\newcommand{\stagenum}[1]{(\textbf{#1})}

\title{Dynamical-Model-Informed Stochastic Flow Matching for Global Probabilistic Subseasonal Forecasting of Means and Extremes}
\author{
Abdullah A. Fahad$^{1,*}$, Manmeet Singh$^{2}$\\[0.5em]
\footnotesize $^{1}$Global Modeling and Assimilation Office, National Aeronautics and Space Administration (NASA) Goddard Space Flight Center,\\
\footnotesize Greenbelt, Maryland 20771, United States\\[0.3em]
\footnotesize $^{2}$Department of Earth, Environmental and Atmospheric Sciences, Western Kentucky University,\\
\footnotesize Bowling Green, Kentucky 42101, United States\\[0.5em]
\footnotesize $^{*}$Corresponding author: \href{mailto:a.fahad@nasa.gov}{a.fahad@nasa.gov}
}
\date{}

\begin{document}
\maketitle

\begin{abstract}
Physics-based models produce skillful subseasonal forecasts but suffer from systematic biases, weakened teleconnections, and parameterized physics, and ensembles large enough for probabilistic detection of extremes are computationally expensive; purely data-driven models see too few observed realizations of the low-frequency modes that carry subseasonal predictability. We introduce \methodname{}, a hybrid framework that turns \modelshort{} subseasonal ensemble guidance into 90-member stochastic forecasts of weekly precipitation and 2\,m temperature on a global $1^\circ$ grid via a lead-aware flow-matching U-Net. Its physically informed stochastic prior initializes members from empirical orthogonal function (EOF) perturbations conditioned on the Madden--Julian Oscillation, North Atlantic Oscillation, El Ni\~no--Southern Oscillation, and lead week rather than white noise, so the learned flow refines large-scale covariance rather than inventing it. On held-out 2021--2025 forecasts, the standard empirical CRPS in a same-initialization four-versus-four comparison improves by 22.0\% (precipitation) and 25.6\% (temperature) in the lead mean; the corresponding 90-versus-four production comparison improves by 32.2\% and 34.0\%. Equal member count controls the direct member-count advantage but does not eliminate the distribution-dependent finite-ensemble bias of empirical CRPS. A two-factor ablation over five training seeds attributes distinct gains to the structured prior and a learned error-variance pathway, the prior being the larger contributor. Conditional verification over 120 observed-extreme cases shows positive CRPS skill, but the ensemble remains underdispersive, four-member upper quantiles are poorly resolved, and unconditional precipitation exceedance BSS is near neutral after week~1. The results therefore support improved probabilistic skill and tail representation, not calibrated operational rare-event probabilities. Case studies of the 2022 United Kingdom heatwave, the 2023 California atmospheric rivers, and a shared miss of the 2022 Pakistan floods show the reach and limits of dynamically informed refinement.
\end{abstract}

\section{Introduction}
\label{sec:intro}

Subseasonal-to-seasonal (S2S) prediction occupies the difficult interval between deterministic weather prediction and seasonal climate outlooks. The week-3/4 range is sometimes described as a predictability ``desert'' because forecast skill there is systematically lower than at either flanking timescale \citep{Vitart2017}. At this range, atmospheric initial conditions lose deterministic control while slower land, ocean, and tropical modes can still provide useful predictability \citep{Vitart2017,RobertsonVitart2018}. Forecast users nevertheless need calibrated probabilities rather than a single best estimate, and for precipitation and temperature extremes the requirement is stricter still: the decision-relevant quantities are exceedance probabilities and upper quantiles, which small dynamical ensembles can neither resolve nor calibrate. This motivates methods that use physically based dynamical guidance while correcting systematic forecast errors, quantifying residual uncertainty, and sampling the forecast tail densely enough for extreme-event probabilities to be estimated at all.

Neither purely dynamical nor purely data-driven prediction resolves this problem alone, and the two approaches fall short in complementary ways. Coupled dynamical systems already deliver useful S2S skill and can anticipate some large-scale extremes weeks in advance, as demonstrated across the multi-model S2S database, including elevated probabilities for the 2010 Russian heatwave up to three weeks ahead \citep{Vitart2017,VitartRobertson2018}. Their limits are practical and structural. Realistic probabilistic forecasts, and especially tail probabilities, require large ensembles, yet every added member costs a full coupled integration, and the burden compounds at the higher resolutions needed to represent extremes \citep{LeutbecherPalmer2008,Bauer2015}; many operational S2S systems therefore run only a handful of members per initialization (four in the archive used here), and even the largest ensembles remain too small to resolve local rare-event probabilities. Dynamical models also carry systematic biases rooted in parameterized physics, and they commonly misrepresent the strength and pathways of the tropical--extratropical teleconnections that carry much of the S2S signal \citep{Stan2017}. Machine-learning forecast systems invert this trade: once trained, they generate members at negligible cost, are not tied to the resolution of a dynamical core, and have reached or surpassed operational medium-range skill in both deterministic and probabilistic settings \citep{Bi2023,Lam2023,Price2025}. But they face their own structural constraint at subseasonal range. The reanalysis and satellite record spans only a few decades, which contains many realizations of synoptic weather but few independent realizations of the low-frequency modes, the Madden--Julian Oscillation (MJO), the North Atlantic Oscillation (NAO), and El Ni\~no--Southern Oscillation (ENSO) cycles, that supply much of the predictability beyond week two, so a purely data-driven model has limited opportunity to learn their evolution and remote impacts from data alone \citep{WatsonParris2021}. These complementary failure modes motivate hybrid designs that retain the dynamical model's physically consistent evolution of the slow components while using machine learning for error correction, uncertainty quantification, and inexpensive ensemble generation \citep{Kochkov2024,Mouatadid2023}; the framework developed here belongs to this family.

\modelshort{} subseasonal forecasts provide a physically grounded ensemble prior but, like other dynamical S2S systems, contain flow-dependent bias, regional calibration errors, under- and overdispersion, and limited precipitation sharpness at longer leads. Statistical postprocessing and ensemble model output statistics have long been used to improve weather forecasts \citep{Gneiting2005,RaspLerch2018,Gronquist2021}; an explicit spatial model is a plausible route to member-level coherence, but that property must be verified with multivariate scores rather than inferred from architecture or maps. Stochastic neural forecasts meanwhile increasingly emphasize ensemble distributions and proper scoring rules rather than deterministic trajectory error alone \citep{Lang2026,Price2025}. For subseasonal impacts, however, a useful system should neither discard dynamical model information nor evaluate rare events only through an ensemble mean.

We frame the problem as probabilistic forecast refinement. \modelshort{} supplies the dynamical forecast state; observed ocean, land, circulation, moisture, and MJO-related predictors supply recent boundary and large-scale context; static geography supplies persistent spatial structure; and a flow-matching model learns an observation-consistent distribution over weekly precipitation and 2\,m temperature. Flow matching trains continuous generative models by regressing velocity fields along probability paths \citep{Lipman2023,Liu2023}, making it a natural fit for stochastic ensemble generation when paired with climate-informed initial perturbations. In this setting the ensemble noise is not a disposable implementation detail: it defines the starting distribution of every forecast member and therefore controls which spatial covariance structures the learned velocity field must refine.

The contributions are fourfold. First, we formulate a global $1^\circ$ weekly S2S forecast-refinement task for precipitation and 2\,m temperature conditioned on \modelshort{} ensemble summaries. Second, we implement a lead-aware multi-target flow-matching architecture with static geography, global-context cross-attention, separate precipitation and temperature heads, and structured velocity losses, reducing competition between short-lead synoptic corrections and longer-lead planetary-scale behavior. Third, we introduce a physically informed stochastic ensemble prior: EOF perturbations conditioned on MJO, NAO, ENSO, and lead week are blended with random support and variance-tempered spread adjustment, and the same structured-noise family is used during training and inference to avoid prior mismatch. Fourth, we evaluate routine skill, the stochastic prior, ensemble-generation properties, and high-impact extremes as related but distinct verification problems, using CRPS, root-mean-square error (RMSE), bias, correlation, member-count convergence diagnostics, Brier skill scores (BSS), season--lead matrices, spatial skill maps, a 120-case extreme-event catalog (60 precipitation, 60 temperature), and named-event tail diagnostics. This evaluation includes a documented shared miss that motivates further study of difficult high-impact precipitation events.

\begin{figure}[!tbp]
\centering
\includegraphics[width=0.9\linewidth]{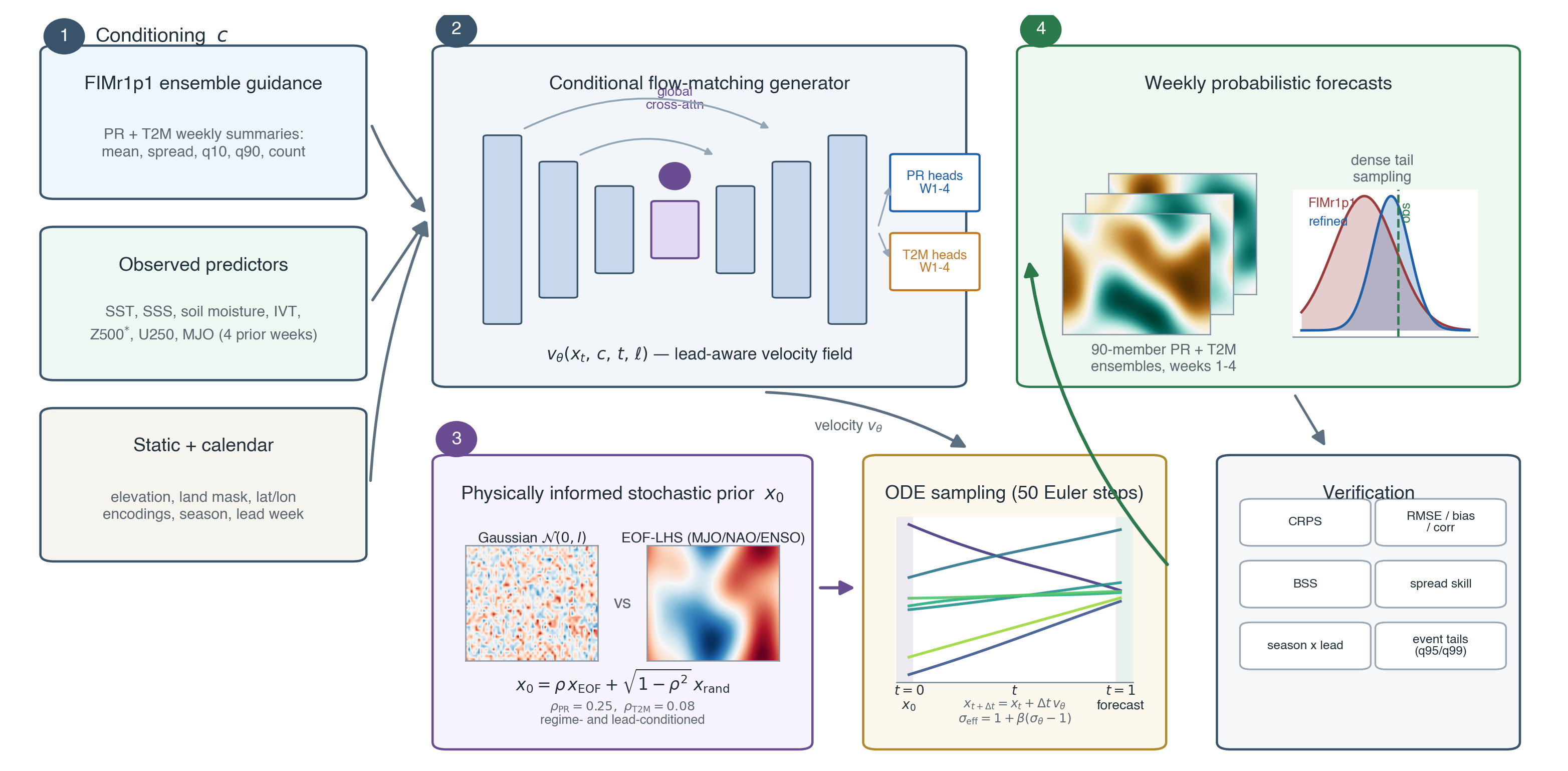}
\caption{The \methodname{} forecast-refinement framework. \stagenum{1} \modelshort{} ensemble summaries, recent observed predictors, and static/calendar fields form the conditioning tensor $c$. \stagenum{2} A lead-aware flow-matching U-Net with global-context cross-attention and separate precipitation/temperature heads learns the conditional velocity field $v_\theta(x_t,c,t,\ell)$. \stagenum{3} Ensemble members are initialized from a physically informed stochastic prior, built from regime-conditioned EOF-LHS perturbations blended with Gaussian support ($x_0=\rho\,x_{\mathrm{EOF}}+\sqrt{1-\rho^2}\,x_{\mathrm{rand}}$), and integrated to $t{=}1$ with 50 Euler steps under variance-tempered spread control. \stagenum{4} The result is a 90-member weekly precipitation and 2\,m temperature ensemble verified with all-case, spatial, calibration, and event-tail diagnostics. The two rendered noise fields contrast the isotropic Gaussian prior with the spatially coherent EOF-LHS prior whose contribution is examined by the prior ablation in Sect.~\ref{sec:results-noise}.}
\label{fig:overview}
\end{figure}

\section{Data and Forecast Problem}
\label{sec:data}

\subsection{Prediction task}

The prediction task is global weekly subseasonal forecasting on a $181\times360$ latitude--longitude grid. For each weekly \modelshort{} initialization the system forecasts four weekly leads; lead week $\ell\in\{1,\dots,4\}$ is the 7-day mean over forecast days $7(\ell-1)$ through $7\ell-1$, with day~0 the initialization day, so week~1 covers the initialization day and the following six days and the four leads span forecast days 0--27. The evaluation period contains 52 initializations per year. The raw \modelshort{} archive provides four ensemble members per initialization, whereas the neural system generates 90. Because empirical CRPS favors larger ensembles at finite member count \citep{Ferro2014}, the central model-quality comparison uses four generated members against the four raw members of the identical initialization (Sect.~4.3). We call this comparison \emph{ensemble-size matched}, not unbiased: matching $M$ removes the obvious 90-versus-four sampling advantage, but the remaining empirical-CRPS bias can differ between forecast distributions with different spread. Training and inference condition on fixed summary statistics of the four archive members. This interface can accept the same statistics computed from another member count, but performance at a dynamical ensemble size other than four has not been demonstrated; we therefore avoid claiming empirical ensemble-size agnosticism. The full production configuration, the 90-member generated ensemble against the raw four-member archive, is reported alongside because inexpensive ensemble enlargement is part of the system's design. The target variables are weekly mean precipitation (PR), expressed as the weekly mean daily rate in mm\,day$^{-1}$, and weekly mean 2\,m temperature (T2M). PR is verified against precipitation from the daily Global Precipitation Climatology Project (GPCP) Climate Data Record (CDR), version 1.3 \citep{GPCPdaily2017,Huffman2001}, whose native $1^\circ$ daily analyses are aligned to the forecast grid by bilinear interpolation and averaged to weekly means over the target windows defined above; T2M is verified against temperature derived from the fifth-generation European Centre for Medium-Range Weather Forecasts (ECMWF) reanalysis (ERA5) \citep{Hersbach2020}. PR is predicted in physical units after a square-root target transform. T2M is represented during training as the residual between observed ERA5 T2M and the target-lead \modelshort{} ensemble-mean T2M; the residual is added back to \modelshort{} at inference.

\paragraph{Data splits and tuning protocol.} Training uses 1999--2019, and 2020 is the development year. All architectural and sampling hyperparameters were selected on 2020 validation scores only; this includes the prior blend weights $\rho_q$, the tempering coefficients $\beta_q$, the lead-loss weights, and the three-stage training schedule of Sect.~4.1, whose stage transitions and early stopping are driven by 2020 validation scores. The 2021--2025 evaluation period was never consulted for any design, tuning, or model-selection decision. All results are computed from a single frozen checkpoint and a frozen generated-forecast archive.

\subsection{Dynamical and observed predictors}

The model conditions on weekly ensemble information from the \modelname{} (\modelshort{}) for PR and T2M. \modelshort{} is included in the Subseasonal Experiment (SubX) archive \citep{Pegion2019SubX}, is coupled with ocean, land, and sea-ice components, and is documented specifically for subseasonal forecasting by \citet{Sun2018FIM}. Rather than passing only an ensemble mean, the system ingests five \modelshort{} summary statistics for each variable and lead: mean, standard deviation, 10th percentile, 90th percentile, and member count. All sample quantiles in conditioning and verification use linear type-7 interpolation; with four members, q10 and q90 are therefore interpolated near the minimum and maximum. The archive provides four ensemble members per initialization, each extending 32 daily leads, throughout the training and evaluation years. The archive labels its daily fields lead days 1--32, with lead day 1 valid on the initialization day; forecast weekly means are computed from the first 28 archived daily fields in consecutive 7-day blocks, so archive lead days 1--28 correspond exactly to forecast days 0--27 in the day-0 convention of Sect.~2.1, and the forecast and observed weekly means cover identical calendar windows. The summary statistics are therefore computed from four members at both training and inference; the neural system never conditions on more dynamical information than the raw four-member forecast of its own initialization. The interface is fixed-dimensional and includes member count, but member count is constant at four in all experiments; generalization to another dynamical ensemble size is untested. Only the statistics for the target lead are used, which keeps the conditioning compact and aligns the correction with the verification lead. Predictor and target normalization statistics are frozen from 1999--2019 only.

Observed predictors are grouped into local target-grid predictors and global context predictors, each represented as four consecutive 7-day means ending the day before initialization, covering days $t_0-28$ through $t_0-1$. These are \emph{valid-time} cutoffs, not an audit of historical publication latency: no predictor window uses a valid time after initialization, but the final retrospective products used here may not have been published by the nominal issue date. Local predictors are soil moisture and an MJO wave field; global context predictors are sea-surface temperature (SST), sea-surface salinity (SSS), integrated vapor transport (IVT), 500\,hPa geopotential-height zonal deviation (Z500), and 250\,hPa zonal wind (U250). Each predictor targets a specific, physically motivated source of multiweek predictability. SST anomalies carry the slow oceanic boundary forcing, ENSO foremost among the associated modes, that organizes tropical convection and its extratropical teleconnections at weeks 2--4 \citep{Stan2017}. SSS complements SST as a tracer of upper-ocean freshwater stratification and of the ocean's imprint of moisture export, and near-surface salinity has been shown to carry predictive information for downstream terrestrial rainfall \citep{Li2016SSS}. IVT marks active moisture corridors, atmospheric rivers and monsoon moisture surges, that supply heavy-precipitation events \citep{ZhuNewell1998}. The Z500 zonal deviation isolates quasi-stationary waves and blocking, the persistent circulation regimes behind multiweek temperature extremes, and U250 tracks the position and strength of the upper-level jet that acts as the Rossby waveguide steering those teleconnections \citep{Stan2017}. On the land side, soil moisture supplies weeks-to-months memory and couples to temperature extremes through evaporative regimes, a leading terrestrial source of subseasonal predictability \citep{Koster2004}, and the MJO wave field summarizes the dominant intraseasonal tropical mode whose global teleconnections modulate both precipitation and temperature across all four target leads \citep{WheelerHendon2004}. Providing four consecutive pre-initialization weeks exposes both recent state and tendency. Data sources are the NOAA Optimum Interpolation Sea Surface Temperature (OISST) product for SST \citep{Reynolds2007}, Copernicus SSS \citep{CopernicusMarineSSS}, Copernicus Climate Change Service (C3S)/European Space Agency (ESA) soil moisture \citep{Dorigo2017,C3SSoilMoisture}, and ERA5-derived circulation and moisture fields \citep{Hersbach2020}. Table~\ref{tab:data-config} records the valid-time cutoff; a point-in-time availability archive was not used for any of these sources or for the climate indices. The system is therefore a research hindcast postprocessor, not a demonstrated operational pipeline (Sect.~\ref{sec:limitations}).

\subsection{Static geography}

Five static channels provide persistent spatial context: min--max-normalized elevation derived from Global 30 Arc-Second Elevation (GTOPO30), a land mask, normalized latitude, and sine/cosine longitude. These allow the neural correction to depend on coastlines, orography, land--ocean contrast, and periodic longitude structure without relearning them from dynamic predictors. Table~\ref{tab:data-config} summarizes the full configuration.

\begin{table}[!tbp]
\centering
\caption{Data, model, and training configuration.}
\label{tab:data-config}
\small
\begin{tabular}{>{\raggedright\arraybackslash}p{0.30\linewidth}>{\raggedright\arraybackslash}p{0.62\linewidth}}
\toprule
Item & Configuration \\
\midrule
Target grid & Global $181\times360$, $1^\circ$ resolution \\
Forecast leads & Four weekly leads, weeks 1--4 \\
Targets & GPCP PR; ERA5 T2M residual relative to \modelshort{} ensemble mean \\
Dynamical input & \modelshort{} PR and T2M ensemble summaries: mean, standard deviation, q10, q90, member count \\
Observed local predictors & Soil moisture and MJO wave fields, four pre-initialization weeks \\
Observed global context & SST, SSS, IVT, Z500 zonal deviation, U250, four pre-initialization weeks \\
Static inputs & Elevation, land mask, normalized latitude, sine/cosine longitude \\
Training years & 1999--2019 \\
Validation year & 2020 (all tuning and model selection) \\
Evaluation years & 2021--2025 (untouched during development) \\
Test ensemble & 90 generated members, 50 Euler steps \\
Skill evaluation ensembles & Production: 90 generated vs raw 4-member archive; size-matched comparison: repeated 4-member generated subsets vs the 4 raw members of the same initialization (Sect.~4.3) \\
Ensemble prior & MJO/NAO/ENSO- and lead-conditioned EOF-LHS perturbations blended with Gaussian support; variance-tempered spread at inference \\
\midrule
Velocity backbone & U-Net, 4 resolution levels, channels 128/256/512/768, 2 residual layers per block, GroupNorm/SiLU \\
Global-context pathway & Convolutional encoder to 128-dim spatial tokens; bottleneck cross-attention, 4 heads, 1 layer \\
Conditioning encoders & Context MLP (hidden 128) on spatial summary statistics; learned lead embedding ($4\times128$) \\
Output heads & Per-lead, per-variable mean and variance heads (hidden 64, bottleneck 32) \\
Parameter accounting & Exact trainable count is reported with the archived checkpoint metadata; all widths and modules needed to reproduce the architecture are listed here and in Table~S1 \\
Optimization & Batch size 4, learning rate $2\times10^{-5}$, weight decay $10^{-4}$, BF16; three-stage schedule (Sect.~4.1) \\
Prior/spread settings & $\rho_{\mathrm{PR}}=0.25$, $\rho_{\mathrm{T2M}}=0.08$; $\beta_{\mathrm{PR}}=0.45$, $\beta_{\mathrm{T2M}}=0.03$; $K=30$ EOF modes per category \\
\bottomrule
\end{tabular}
\end{table}

\section{Method}
\label{sec:method}

\subsection{Forecast refinement view}

Let $y_{\ell}$ denote the observed weekly target field at lead week $\ell\in\{1,2,3,4\}$, $g_{\ell}$ the \modelshort{} ensemble information for the same lead, $z$ the recent observed predictors, and $s$ the static geography and calendar information. The goal is not to replace \modelshort{} with a free-running neural forecast but to learn a conditional forecast distribution
\begin{equation}
  p_{\theta}(y_{\ell}\mid c), \qquad c = (g_{\ell}, z, s, \ell),
\end{equation}
where $c$ is the conditioning tensor used throughout; in the flow-matching notation below, $x_1$ denotes the normalized representation of $y_{\ell}$. Here $g_{\ell}$ comprises the five \modelshort{} ensemble summary channels (including the member count) and $s$ comprises the static geography and calendar encodings; Table~\ref{tab:data-config} is the definitive inventory of all conditioning fields. The dynamical model provides a physically informed prior state; machine learning (ML) provides observation-consistent distributional corrections on a spatial grid. Member-level spatial and cross-variable coherence are hypotheses of the architecture, not verified results in the present score suite. An immediate corollary, examined explicitly in Sect.~\ref{sec:results-events}, is that the refinement is conditional on its inputs: it can sharpen and sometimes relocate probability mass relative to the dynamical ensemble, but failures by both systems require event-level diagnosis.

\subsection{Conditional flow matching}

Let $x_1$ be the normalized target field, i.e., the transformed and normalized representation of the observed target $y_\ell$ of Sect.~3.1, and let $x_0$ be a stochastic initial field. For $t\in[0,1]$ the model is trained on the linear path
\begin{equation}
  x_t = t\,x_1 + (1-t)\,x_0,
\qquad
  v^\star = x_1 - x_0 .
\end{equation}
The network predicts $v_{\theta}(x_t,c,t,\ell)$, where $c$ is the conditioning tensor. The base velocity objective is
\begin{equation}
  \mathcal{L}_{\mathrm{vel}} =
  \frac{1}{N}\sum_i w_{\mathrm{area},i}\, w_{\mathrm{lead},i}
  \left\lVert v_{\theta}(x_t,c,t,\ell)_i - v^\star_i \right\rVert_2^2,
\end{equation}
where $w_{\mathrm{area}}$ is a latitude-cosine area weight and $w_{\mathrm{lead}}$ increases from week~1 to week~4 as $(1.0,\,1.2,\,1.5,\,2.0)$, reflecting the greater difficulty and user value of long-lead corrections. The lead weighting was motivated during development by loss balance: long-lead velocity errors have smaller amplitudes, and without up-weighting they are dominated in the loss by short-lead errors. It was not selected to maximize week-3/4 verification skill. Both the lead weights and the flow-time sampling schedule (which emphasizes low and middle interpolation times) are nevertheless design choices adopted without dedicated ablation; we flag them as untested and return to this in Sect.~\ref{sec:limitations}. Forecast samples are generated by explicit Euler integration,
\begin{equation}
  x_{t+\Delta t} = x_t + \Delta t\, v_{\theta}(x_t,c,t,\ell),
\end{equation}
from $t=0$ to $t=1$ in 50 steps; the test configuration draws 90 ensemble members.

\subsection{Architecture}

The velocity model wraps a U-Net backbone \citep{Ronneberger2015} in the flow-matching formulation (Fig.~\ref{fig:overview}, stage~2). The state channels are the precipitation and 2\,m temperature residual/target fields; conditioning channels are concatenated with the flow state, and static geography is appended internally. A context multilayer perceptron (MLP) summarizes the conditioning tensor through spatial means and standard deviations, lead week is represented by a learned embedding, and a global-context encoder maps coarse global predictor fields into spatial tokens. A bottleneck cross-attention module lets local U-Net bottleneck cells attend to these global tokens. Encoder/decoder widths, attention dimensions, output heads, and training/sampling settings are listed in Supplement Table~S1; exact total and trainable parameter counts are retained with the archived checkpoint metadata.

The output stage contains separate lead-specific precipitation and temperature heads, plus separate variance heads for each variable and lead. This shares most spatial representation learning while allowing week-1 and week-4 corrections to specialize, avoiding capacity competition between filters that correct short-lead weather-scale errors and filters that represent slower week-3/4 circulation. It also avoids forcing precipitation and temperature, which have very different spatial and distributional structure, through a single final projection.

\subsection{Target transforms and physical bounds}

Precipitation is square-root transformed before normalization and decoded by squaring, with a nonnegative clamp; this compresses the dynamic range of heavy precipitation and prevents negative rain at inference. Temperature is trained as a \modelshort{} residual bounded by global statistics and added back to the target-lead \modelshort{} ensemble mean at decode time. These transformations are not conservation constraints, but they impose simple physical bounds that keep generated samples meteorologically plausible.

The asymmetric target design, residual for temperature but direct prediction for precipitation, is deliberate and reflects how each variable relates to the dynamical guidance. The \modelshort{} T2M ensemble mean is already highly skillful, and its errors are smooth, close to Gaussian, and regionally systematic; a residual target therefore removes the large climatological range from the learning problem and concentrates model capacity on exactly that error structure, while the decoded forecast inherits the dynamical model's synoptic evolution by construction. Precipitation offers no such well-behaved reference. Weekly rainfall is intermittent, zero-bounded, and strongly non-Gaussian, and \modelshort{} precipitation errors are dominated by spatial displacement of rain features; the residual between an observed and a displaced forecast feature is a dipole with positive and negative lobes, which is harder to model than the rain field itself, breaks the zero bound, and is incompatible with the square-root transform. We verified this reasoning during development: a precipitation-as-residual configuration was tested and degraded 2020 validation CRPS relative to direct prediction, with the largest deterioration for heavy-rain amounts, and was therefore discarded.

\subsection{Structured velocity loss}

The velocity loss is augmented with variable-specific spatial-structure terms: a pixel loss, first-gradient losses in latitude and longitude, and coarse losses after $2\times$ and $4\times$ average pooling,
\begin{equation}
  \mathcal{L}_{q} =
  \mathcal{L}_{\mathrm{pix},q}
  + \lambda_{\nabla,q}\mathcal{L}_{\nabla,q}
  + \lambda_{2,q}\mathcal{L}_{2,q}
  + \lambda_{4,q}\mathcal{L}_{4,q},
  \qquad q\in\{\mathrm{PR},\mathrm{T2M}\},
\end{equation}
with the final loss averaging the two per-variable totals. The configured coefficients are $\lambda_{\nabla,\mathrm{PR}}=0.03$, $\lambda_{\nabla,\mathrm{T2M}}=0.01$, $\lambda_{2,\mathrm{PR}}=0.03$, $\lambda_{2,\mathrm{T2M}}=0.01$, $\lambda_{4,\mathrm{PR}}=0.02$, and $\lambda_{4,\mathrm{T2M}}=0.01$; they were fixed during development on 2020 validation scores, and no systematic sensitivity analysis of these coefficients was performed. This term encourages the learned velocity field to preserve both local gradients and broader spatial organization.

\subsection{Physically informed stochastic prior}
\label{sec:method-prior}

The stochastic forecast distribution begins with the initial field $x_0$. In a conventional generative forecast $x_0$ is drawn from independent Gaussian noise. That choice is mathematically convenient, but it asks the learned velocity field to create all coherent precipitation and temperature structure during the ordinary differential equation (ODE) integration. For subseasonal prediction this is an unnecessarily weak prior, because large-scale modes already constrain plausible spatial covariance. \methodname{} therefore treats the ensemble perturbation itself as a physical input to the forecast system.

EOF perturbations are conditioned on MJO, NAO, ENSO, and lead-week state. The MJO is a known source of subseasonal tropical and extratropical predictability \citep{MaddenJulian1971,WheelerHendon2004}; NAO and ENSO states provide additional large-scale circulation and oceanic regime information \citep{WallaceGutzler1981,BarnstonLivezey1987}; EOF bases compactly represent recurring spatial modes \citep{Lorenz1956}. These indices are used as retrospective covariates in this study. In particular, ``most recent'' below means the most recent value in the retrospective archive consistent with the stated valid-time rule, not a reconstruction from historically published data vintages.

The prior is constructed as follows. EOF bases are computed separately for each target variable: precipitation bases from area-weighted weekly GPCP precipitation anomalies (weekly means of the daily v1.3 CDR; Sect.~2.1) and temperature bases from area-weighted weekly ERA5 2\,m temperature anomalies, in both cases over the 1999--2019 training period only, so no information from the 2020 validation year or the 2021--2025 evaluation years enters the perturbation bases. Anomalies are binned into discrete regime--lead categories before decomposition: nine MJO phases (phase~0 for RMM amplitude $<1$, phases 1--8 otherwise, using RMM indices lagged 28 days before initialization) $\times$ four lead weeks (36 categories); three NAO phases (positive/neutral/negative at $\pm0.5$ on a 7-day trailing mean ending one day before initialization) $\times$ four leads (12 categories); and three ENSO states at ONI $\pm0.5$ $\times$ four leads (12 categories). The 28-day RMM lag is a conservative modeling choice, not a requirement imposed by RMM publication latency; a 28-day-old MJO state is dynamically distinct from a near-initialization state, and no $t_0-1$ sensitivity experiment is reported. Within each category the leading $K=30$ EOFs and eigenvalues are retained; categories with fewer than 20 samples fall back to phase-only bases and then to an all-data basis. At sampling time the regime state uses retrospective RMM, NAO, and ONI archives under these valid-time rules rather than point-in-time published vintages. For each member, variable, and regime family, EOF coefficients are drawn by Latin hypercube sampling with eigenvalue scaling and each sampled field is normalized to unit variance. The three regime components are then combined with an independent Gaussian field $n_{\mathrm{iid}}$ as
\begin{equation}
  x_{\mathrm{EOF}} = \mathcal{S}\!\left[\,0.9\,\bigl(w_{\mathrm{MJO}}\,n_{\mathrm{MJO}} + w_{\mathrm{NAO}}\,n_{\mathrm{NAO}} + w_{\mathrm{ENSO}}\,n_{\mathrm{ENSO}}\bigr) + 0.1\,n_{\mathrm{iid}}\,\right],
\label{eq:eofmix}
\end{equation}
where $\mathcal{S}[\cdot]$ normalizes to unit variance and the weights $w$ are proportional to the concurrent regime amplitudes, i.e., the lagged RMM amplitude clipped to $[0.1,3.0]$ and the absolute values of the lagged NAO index and the ONI each clipped to $[0.1,2.5]$, normalized to sum to one, so stronger concurrent regimes contribute proportionally more structured covariance; the construction is applied independently for each target variable $q$ with that variable's bases, yielding $x_{\mathrm{EOF}}^{(q)}$. For each variable $q$ the structured and random components are blended as
\begin{equation}
  x_0^{(q)} = \rho_q\, x_{\mathrm{EOF}}^{(q)} + \sqrt{1-\rho_q^2}\; x_{\mathrm{rand}}^{(q)},
\label{eq:blend}
\end{equation}
with $\rho_{\mathrm{PR}}=0.25$ and $\rho_{\mathrm{T2M}}=0.08$ at evaluation time (ramped from 0.15 and 0.05 during early training); the evaluation-time values were chosen from a small set of candidates by minimizing 2020 validation ensemble CRPS, separately per variable. Precipitation deliberately receives the stronger structured prior, reflecting the greater importance of organized tropical convection, monsoon structure, and spatial displacement uncertainty for weekly precipitation.

Each channel of $x_0$ therefore draws from its own variable's bases, the precipitation channel from the precipitation-anomaly EOFs and the temperature channel from the temperature-anomaly EOFs, with coefficient draws that are independent between channels, and each structured field is normalized to unit variance before blending; because $x_{\mathrm{EOF}}^{(q)}$ and $x_{\mathrm{rand}}^{(q)}$ are independent with unit variance, the $\sqrt{1-\rho_q^2}$ weight in Eq.~(\ref{eq:blend}) preserves unit variance in $x_0^{(q)}$. Because the coefficient draws are independent between channels, the prior does not represent cross-variable covariance between precipitation and temperature. The structured component is deliberately weighted much more strongly for precipitation: at $\rho_{\mathrm{T2M}}=0.08$ it contributes $\rho^2\approx0.6\%$ of the initial temperature variance, so the temperature prior remains close to Gaussian support, whereas for precipitation ($\rho_{\mathrm{PR}}=0.25$, $\rho^2\approx6\%$) the structured covariance is a material part of the initialization. The naming should also be read precisely: the prior is statistically derived from EOFs of observed anomalies and lagged climate indices, so it injects observed covariance structure and regime dependence into the initial ensemble, but it does not enforce physical laws.

Using structured noise only at inference would expose the ODE solver to initial states the velocity model never saw during training. Training therefore uses the same EOF-LHS mixture, so the learned flow matches the sampling path used at evaluation: the initial state already contains broad spatial covariance, and the velocity field can focus on observation-conditioned refinement rather than constructing planetary-scale structure from white noise. The decisive test is not ``more ensemble members'' but whether a model trained and sampled with the physically informed prior improves CRPS and tail probabilities relative to a counterpart model trained and sampled with Gaussian noise, at the same ensemble size and ODE-step count; that controlled comparison is reported in Sect.~\ref{sec:results-noise}.

\subsection{Variance-tempered spread control}

Variance heads estimate a relative spread multiplier $\sigma_\theta$, trained first against a normalized absolute velocity residual and regularized toward a physically moderate range, then fine-tuned directly with an ensemble CRPS objective in the final training stage (Sect.~4.1). At inference the multiplier is clipped, coarse-grained over an 8-grid-cell kernel, and tempered toward one,
\begin{equation}
  \sigma_{\mathrm{eff}} = 1 + \beta_q(\sigma_{\theta}-1),
\end{equation}
with $\beta_{\mathrm{PR}}=0.45$ and $\beta_{\mathrm{T2M}}=0.03$, likewise chosen from small candidate sets by 2020 validation ensemble CRPS, separately per variable. This lets the ensemble express heteroscedastic uncertainty without allowing the variance pathway to dominate the flow trajectory; the pathway's skill contribution is isolated by the two-factor ablation of Sect.~\ref{sec:results-noise}. Note that $\beta_{\mathrm{T2M}}=0.03$ tempers the temperature multiplier nearly to unity, so the variance pathway is effectively disabled for temperature and T2M ensemble spread is governed almost entirely by the stochastic prior and the learned flow. This setting was selected on 2020 validation, where stronger temperature tempering degraded calibration; the variance head remains most useful for precipitation, whose uncertainty is strongly heteroscedastic.

\FloatBarrier

\section{Experimental Design}
\label{sec:design}

\subsection{Training configuration}

The production configuration trains from scratch with batch size 4, learning rate $2\times10^{-5}$, weight decay $10^{-4}$, and bfloat16 (BF16) mixed precision, in three stages. Stage~1 trains the velocity pathway alone for 25 epochs, stabilizing the learned flow before any variance learning is introduced. Stage~2 mixes the variance-head loss into the total objective with weight 0.3 and continues joint training until the validation loss saturates, up to a maximum of 100 epochs; the best checkpoint by 2020 validation score is retained. Stage~3 fine-tunes the uncertainty representation only: starting from that best checkpoint, the velocity U-Net is frozen, the optimizer state is reset, and the variance heads are trained directly with an ensemble CRPS objective computed from small generated ensembles, with early stopping once the 2020 validation CRPS fails to improve for 15 consecutive epochs. Separating stage~3 in this way lets the spread representation be optimized against a proper probabilistic score without perturbing the converged velocity field. Context dropout is applied to non-\modelshort{} predictors during training, but \modelshort{} conditioning is always kept active because it is the operational dynamical prior. The flow-time sampler emphasizes low and middle interpolation times, increasing exposure to the early portion of the ODE trajectory where the stochastic initial condition most strongly shapes ensemble structure.

\paragraph{Training sample size.} The training record is capped by the availability of the dynamical hindcasts themselves: the \modelshort{} archive begins in 1999, so 21 years of weekly initializations are available for training, roughly 1{,}100 forecasts and about 4{,}400 lead-specific global forecast--observation field pairs. Grid cells and leads provide many loss terms, and redrawing $x_0$ and $t$ provides stochastic augmentation of the regression objective, but neither operation creates additional independent meteorological realizations: spatial cells, leads from the same initialization, and neighboring weekly starts are correlated. The number of independent low-frequency-regime realizations therefore remains limited by the 21-year record. Conditional refinement around dynamical guidance and the structured prior reduce the complexity of the learning problem, but they do not remove this sample-size limitation. Consistent with that qualification, the five-seed ablation shows between-seed 95\% intervals of $\pm1.3$--4.7 skill points (Sect.~\ref{sec:results-noise}), materially larger than the temporal bootstrap intervals of a single checkpoint.

\subsection{Baselines and reference configurations}

The dynamical baseline is \modelshort{} on the same variable, lead, initialization, and grid, because \modelshort{} is both the conditioning source and the forecast being refined. Five reported ensemble configurations are used, and every table and figure states which one produced it. (i)~\emph{Same-initialization size-matched comparison} (Tables~\ref{tab:fair-eight-results} and \ref{tab:extreme-event-results}, Fig.~\ref{fig:extreme-skill}): repeated random four-member subsets of the generated archive are scored against the four raw \modelshort{} members of the identical initialization. This controls member count and issue-date information, but it does not make the empirical-CRPS bias identical across distributions. (ii)~\emph{Production configuration} (Tables~\ref{tab:headline-results}--\ref{tab:lead-mask-results}, Figs.~\ref{fig:pr-skill}--\ref{fig:t2m-skill}): the full 90-member generated ensemble against the raw four-member archive ensemble, measuring the delivered research product including denser sampling. (iii)~\emph{Reference checks} (Sects.~\ref{sec:results-overall} and \ref{sec:results-extremes}): a frozen 1999--2019 observed weekly/day-of-year climatological distribution, a lead/month mean-debiased \modelshort{} baseline, and a quantile-mapping (QM) baseline. QM is a standard distributional bias correction for dynamical model output \citep{Cannon2015,Maraun2016}: for each variable, lead week, and grid cell, an empirical transfer function is fit on the 1999--2019 forecast--observation pairs and then applied member-by-member to the four raw members of each 2021--2025 initialization (QM-4). It corrects the marginal forecast distribution but is state-independent and cannot enlarge the ensemble. QM is evaluated here only for the reported week-3/4 extreme cases; this is a focused tail baseline, not an all-case demonstration that a generative architecture dominates every conventional postprocessor. We therefore make no state-of-the-art claim relative to ECMWF or other operational S2S systems. (iv)~\emph{Noise-prior ablation} (Fig.~\ref{fig:noise-ablation}): all four arms of the two-factor design (Sect.~4.4) are scored at 90 members against the same raw four-member \modelshort{} reference. (v)~\emph{Ensemble-as-object diagnostics}: the member-count sweeps (Figs.~\ref{fig:member-convergence}--\ref{fig:member-corr-bss}) and the event case studies (Figs.~\ref{fig:event-pr}--\ref{fig:event-t2m}), which use the full ensemble where tail sampling itself is the object of study.

\subsection{Verification protocol}

The primary all-case probabilistic metric is CRPS \citep{MathesonWinkler1976,Hersbach2000,GneitingRaftery2007}. For an empirical ensemble $\{x_m\}_{m=1}^{M}$ and observation $y$,
\begin{equation}
  \mathrm{CRPS} =
  \frac{1}{M}\sum_{m=1}^{M}\lvert x_m-y\rvert
  - \frac{1}{2M^2}\sum_{m=1}^{M}\sum_{n=1}^{M}\lvert x_m-x_n\rvert ,
\end{equation}
where $m$ and $n$ both index members of the same ensemble.
The size-matched comparison scores repeated random four-member subsets of the 90 generated members against the four raw \modelshort{} members of the identical initialization and averages over subset draws. Matching $M$ controls the direct member-count advantage and holds issue-date information fixed, but it does not remove finite-$M$ bias from the standard empirical estimator because that bias depends on the forecast distribution, particularly its spread \citep{Ferro2014}. Accordingly, we do not describe the four-versus-four score as unbiased. As a distribution-sensitive check we define the fair CRPS estimator
\begin{equation}
  \mathrm{CRPS}_{\mathrm{fair}} =
  \frac{1}{M}\sum_{m=1}^{M}|x_m-y|
  -\frac{1}{2M(M-1)}\sum_{m=1}^{M}\sum_{n\ne m}|x_m-x_n|,
\label{eq:fair-crps}
\end{equation}
which replaces the $M^2$ denominator of the pairwise term by $M(M-1)$. The numerical tables retain the standard empirical CRPS for continuity with the production/member-convergence analysis, so their matched values should be interpreted as equal-$M$ comparisons rather than unbiased distributional estimates. The production configuration (90 generated members versus raw four-member \modelshort{}) is reported separately and must not be interpreted as a pure model-quality comparison. The system-independent climatology reference is frozen from 1999--2019 observations using the same weekly windows and a $\pm2$-week day-of-year pool, so no 2021--2025 observation enters its construction. The QM transfer function $x\mapsto \hat{F}_{\mathrm{obs}}^{-1}(\hat{F}_{\mathrm{fcst}}(x))$ is likewise fit on 1999--2019 and frozen; edge corrections are held constant outside the fitted range and precipitation remains nonnegative.

Deterministic metrics are ensemble-mean RMSE, mean absolute error (MAE), bias, and Pearson correlation. For T2M, whose total-field correlation is dominated by climatology, we additionally report the anomaly correlation (ACC), with anomalies computed relative to an observed weekly climatology built from the 1999--2019 training period. All spatial reductions use latitude-cosine weights, and results are reported both on all grid points (``All'') and on a land mask (``Land''). Season labels, December--February (DJF), March--May (MAM), June--August (JJA), and September--November (SON), always refer to the meteorological season of the verifying week, so a late-December initialization contributes its week-4 lead to DJF of the following year; all 16 season--lead cells are populated. Relative skill for a negatively oriented score $S$ is
\begin{equation}
  \mathrm{Skill}_{S} = 100\left(1 - \frac{S_{\mathrm{ML}}}{S_{\mathrm{raw}}}\right),
\label{eq:skill}
\end{equation}
where $\mathrm{ML}$ denotes \methodname{} and $\mathrm{raw}$ denotes the \modelshort{} reference.
Uncertainty on aggregate scores is quantified with a paired moving-block bootstrap over initialization dates, using blocks of four consecutive weekly starts and 1{,}000 replicates; the same resampled dates are applied to both systems, and 95\% intervals are the 2.5th--97.5th percentiles of the resampled score distribution. Because each initialization enters as one globally area-weighted score, these intervals quantify temporal resampling uncertainty of the aggregate only. They exclude model-training variability, which is quantified separately in the prior ablation: five independently trained checkpoints per configuration give between-seed 95\% intervals between roughly $\pm1.3$ and $\pm4.7$ skill points (Sect.~\ref{sec:results-noise}), considerably wider than the single-checkpoint temporal intervals. Spatial sign-consistency stippling is descriptive, not a statistical significance test. Event-level uncertainty is quantified by paired resampling over cases.

Extremes are evaluated as forecast-distribution problems rather than ensemble-mean reconstruction problems. Events are defined by observed monthly climatological threshold maps, with the default event $y \ge q_{0.95}$ at each grid cell and a minimum precipitation threshold of 5\,mm\,day$^{-1}$; the floor prevents trivially small ``extremes'' in arid regions and dry seasons. For forecast probability $p$ and binary outcome $o$, $\mathrm{BS}=\langle(p-o)^2\rangle$ and $\mathrm{BSS}=1-\mathrm{BS}/\mathrm{BS}_{\mathrm{clim}}$, where the reference probability is the frozen 1999--2019 observed event frequency for the same grid cell and calendar month. Thus no evaluation-year outcome is used to fit the BSS reference. Because BSS is highly sensitive to event base rate and probability reliability, we use CRPS and RMSE as the headline all-case metrics and treat BSS primarily as an event-tail diagnostic. The forecast upper tail is additionally scored directly with the quantile score (QS; pinball loss) at the 95th percentile \citep{GneitingRaftery2007}: for a forecast 95th-percentile field $q$ and observation $y$,
\begin{equation}
  \mathrm{QS}_{0.95}(q,y) =
  \begin{cases}
    0.95\,(y-q), & y \ge q,\\
    0.05\,(q-y), & y < q,
  \end{cases}
\label{eq:pinball}
\end{equation}
so underprediction of an extreme observation is penalized nineteen times more strongly than overprediction. Sample quantiles use linear interpolation between adjacent order statistics (the type-7 convention). For $M=4$, $q_{0.95}$ lies 85\% of the way from the third order statistic to the ensemble maximum and is therefore intrinsically maximum-dominated and convention-sensitive. We consequently treat four-member QS as a secondary diagnostic, emphasize CRPS for the equal-$M$ comparison, and reserve distributional interpretation of the upper tail for the 90-member ensemble. The case-based extreme scores of Sect.~\ref{sec:results-extremes} are conditional verification, measuring performance given that an observed extreme occurred; the all-date exceedance Brier diagnostics are the unconditional assessment of rare-event forecasting, and the two are labeled accordingly.

Ensemble-generation properties are evaluated separately: member-count subsampling experiments on the anomaly-selected extreme-event cases quantify how CRPS and RMSE skill converge with the number of generated members at all four lead weeks, with member-resampling uncertainty bands from repeated random subsets; these bands quantify member-sampling variability conditional on the single frozen generated archive, and do not include uncertainty from training seeds, evaluation-period sampling, or the EOF basis. These convergence diagnostics use the case selection, not the $q_{0.95}$ exceedance masks; the $q_{0.95}$ definition enters only the Brier-score panels. The extreme cases are the deliberate choice here, because rare-event probabilities are exactly the quantities that demand a well-sampled forecast distribution.

\subsection{Noise-prior ablation protocol}

Because \methodname{} generates its ensemble from a stochastic initialization modulated by a learned error-variance pathway, both mechanisms must be evaluated as first-order algorithmic choices. We compare four models in a two-factor design, crossing the ensemble prior (Gaussian $x_0\sim\mathcal{N}(0,I)$ versus regime-conditioned EOF-LHS, Sect.~3.6) with the learned error-variance pathway (disabled versus active, Sect.~3.7), at identical architecture, conditioning fields, data split, training schedule, ensemble size, and ODE solver. Each arm is trained and sampled under its own configuration, so no train--inference mismatch enters any comparison, and comparing the two priors at a fixed variance setting isolates the prior's covariance structure with spread control held constant. Each arm is replicated with five independently trained checkpoints, so between-seed training variability is quantified directly rather than assumed away. The primary ablation score is CRPS skill relative to raw \modelshort{} (Eq.~\ref{eq:skill}); a secondary score is the relative CRPS reduction versus the Gaussian baseline, with architecture and dynamical conditioning held fixed.

\FloatBarrier

\section{Results}
\label{sec:results}

All numerical results are computed from the frozen 2021--2025 evaluation archive, and two comparison configurations run through this section; every table and figure states which one it uses. The \emph{same-initialization size-matched comparison} (Table~\ref{tab:fair-eight-results}) scores repeated four-member generated subsets against the four raw \modelshort{} members of the same initialization. Member counts and issue-date information are equal, but finite-ensemble CRPS bias can still differ between the two distributions (Sect.~4.3). The extreme-event evaluation (Table~\ref{tab:extreme-event-results}, Fig.~\ref{fig:extreme-skill}) uses the same protocol, and the member sweeps (Figs.~\ref{fig:member-convergence}--\ref{fig:member-corr-bss}) scan generated ensemble size against the same raw four-member reference. The \emph{production configuration} (Tables~\ref{tab:headline-results}--\ref{tab:lead-mask-results}, Figs.~\ref{fig:pr-skill}--\ref{fig:t2m-skill}) scores the full 90-member generated ensemble against the raw four-member archive and therefore combines model-distribution differences with a member-count effect. ``All'' rows evaluate all grid points; ``Land'' rows use the land mask. The extreme-event case scores in Sect.~\ref{sec:results-extremes} use 120 cases, 60 precipitation and 60 2\,m temperature, drawn from all regions during 2021--2025. These are conditional scores given an observed extreme. Monthly local observed $q_{0.95}$ threshold maps, with precipitation additionally required to exceed 5\,mm\,day$^{-1}$, define exceedance events only for the unconditional Brier-score diagnostics.

\subsection{Overall skill}
\label{sec:results-overall}

The central equal-$M$ result is the same-initialization size-matched comparison of Table~\ref{tab:fair-eight-results}. With repeated four-member generated subsets scored against the four raw \modelshort{} members of the identical initialization, standard empirical-CRPS skill is positive for every lead, variable, and mask. Lead-mean all-grid skill is 22.0\% for precipitation and 25.6\% for temperature in empirical CRPS, with lead-mean RMSE skill of 19.3\% and 17.1\%; over land the corresponding values are 21.6\% and 23.9\% (CRPS) and 18.9\% and 13.9\% (RMSE). These values control the direct advantage from unequal $M$, but, because the ensembles have different spread, they are not claimed to remove distribution-dependent finite-$M$ CRPS bias.

Table~\ref{tab:headline-results} then reports the production configuration, the forecast product as it would actually be issued: the full 90-member generated ensemble against the raw four-member archive. \methodname{} improves both scores (CRPS and RMSE) for both variables and both masks: all-grid CRPS falls by 32.2\% for precipitation and 34.0\% for temperature, RMSE by 25.6\% and 22.4\%, and land-mask CRPS by 32.3\% and 32.7\%. Among the secondary diagnostics, absolute bias is reduced strongly for precipitation (0.44 to 0.19\,mm\,day$^{-1}$ all-grid) and for land temperature ($-0.31$ to 0.13\,K); the one degraded secondary diagnostic is all-grid temperature bias, which grows from a very small 0.04\,K to a still-small 0.14\,K. We consider this an acceptable trade against the large CRPS and RMSE gains, but report it explicitly. Precipitation pattern correlation increases from 0.576 to 0.711 all-grid (0.603 to 0.745 over land); for temperature, total-field correlations are near unity for both systems and dominated by climatology, so anomaly-based verification (ACC; Sect.~\ref{sec:design}) is the informative deterministic diagnostic and is reported with the robustness checks below. All-case BSS gains are positive throughout but modest; we therefore treat BSS as an event-tail diagnostic (Sect.~\ref{sec:results-extremes}) and rely on CRPS and RMSE for headline all-case skill.

\paragraph{The ensemble-size dividend.}
Moving from four generated members to the 90-member production configuration raises the reported empirical-CRPS skill by roughly 8--11 points (precipitation $22.0\rightarrow32.2\%$, temperature $25.6\rightarrow34.0\%$ all-grid; $21.6\rightarrow32.3\%$ and $23.9\rightarrow32.7\%$ over land) and RMSE skill by 5--7 points. Because both comparisons share the same raw four-member reference, the change measures the combined effect of denser sampling of the generated distribution and the finite-$M$ behavior of the empirical score. We therefore call it a \emph{member-count increment}, not a pure unbiased estimate of model-quality gain. The practical point remains that adding generated members requires only additional learned-flow integrations, whereas adding dynamical members requires coupled-model integrations.

\paragraph{Reference and calibration checks.}
Three checks place both configurations in context. First, anomaly correlation relative to the frozen 1999--2019 observed weekly climatology is 0.46 versus 0.33 (PR) and 0.53 versus 0.44 (T2M) in favor of \methodname{}, showing that the total-field result is not solely a climatological-mean artifact. Second, simple mean debiasing recovers an appreciable share of the T2M improvement, so the temperature result is partly conventional bias correction; the week-3/4 extreme-case QM comparison in Sect.~\ref{sec:results-extremes} is a narrower test of added tail value and is not presented as an all-case baseline. Third, the full 90-member spread/RMSE ratios are 0.80 (PR) and 0.70 (T2M), indicating underdispersion. ERA5 supplies both several predictors and the T2M verification truth, so temperature gains should not be assumed to transfer unchanged to independent observations. We do not use the earlier four-evaluation-year LOYO climatology as evidence because it is a small, noisy reference; the frozen training-period climatology defined in Sect.~4.3 is the appropriate out-of-sample benchmark.

\begin{table}[!tbp]
\centering
\caption{All-case same-initialization, ensemble-size-matched four-versus-four comparison on held-out 2021--2025. Repeated random four-member generated subsets are scored against the four raw \modelshort{} members of the identical initialization (Sect.~4.3). Values use standard empirical CRPS and therefore control member count without eliminating distribution-dependent finite-$M$ bias. The \modelshort{} columns give the raw score (mm\,day$^{-1}$ for PR, K for T2M). Skill entries are percentages with 5th--95th percentile (90\%) temporal bootstrap intervals in parentheses; these intervals exclude training-seed uncertainty.}
\label{tab:fair-eight-results}
\small
\begin{tabular}{lllcccc}
\toprule
Mask & Variable & Lead & \modelshort{} CRPS & CRPS skill (\%) & \modelshort{} RMSE & RMSE skill (\%) \\
\midrule
All & PR & Week 1 & 1.45 & 29.6 (29.2--29.9) & 3.12 & 25.7 (25.3--26.1) \\
All & PR & Week 2 & 1.59 & 20.6 (20.0--21.0) & 3.48 & 18.3 (17.6--18.8) \\
All & PR & Week 3 & 1.64 & 19.4 (18.9--19.7) & 3.57 & 17.6 (17.0--18.0) \\
All & PR & Week 4 & 1.63 & 18.4 (17.8--18.9) & 3.56 & 15.5 (14.8--16.2) \\
All & T2M & Week 1 & 1.01 & 33.4 (32.1--34.5) & 1.76 & 19.2 (18.5--20.1) \\
All & T2M & Week 2 & 1.23 & 25.7 (24.6--26.5) & 2.34 & 16.2 (15.3--17.3) \\
All & T2M & Week 3 & 1.30 & 24.6 (23.7--25.7) & 2.53 & 17.1 (15.8--18.3) \\
All & T2M & Week 4 & 1.34 & 18.5 (17.1--20.0) & 2.61 & 15.9 (14.5--17.7) \\
Land & PR & Week 1 & 1.22 & 29.0 (28.6--29.5) & 2.87 & 24.6 (24.0--25.4) \\
Land & PR & Week 2 & 1.30 & 21.8 (21.3--22.4) & 3.08 & 19.5 (18.7--20.2) \\
Land & PR & Week 3 & 1.32 & 18.6 (17.8--19.0) & 3.14 & 16.9 (16.2--17.6) \\
Land & PR & Week 4 & 1.26 & 17.0 (16.3--17.6) & 3.02 & 14.4 (13.3--15.5) \\
Land & T2M & Week 1 & 1.57 & 31.6 (30.9--32.6) & 2.51 & 17.2 (16.4--18.3) \\
Land & T2M & Week 2 & 2.00 & 22.8 (21.8--23.6) & 3.39 & 11.9 (10.7--13.1) \\
Land & T2M & Week 3 & 2.12 & 21.0 (19.9--22.1) & 3.66 & 12.7 (11.2--14.0) \\
Land & T2M & Week 4 & 2.20 & 20.3 (19.0--21.5) & 3.78 & 14.0 (12.3--15.4) \\
\bottomrule
\end{tabular}
\end{table}

\begin{table}[!tbp]
\centering
\caption{Production-configuration all-case skill on held-out 2021--2025: the full 90-member generated ensemble versus the raw four-member \modelshort{} archive. This comparison combines forecast-distribution differences with a member-count effect and is not an equal-$M$ model-quality estimate. CRPS and RMSE are negatively oriented ($\downarrow$: lower is better); positive skill denotes improvement. Units are mm\,day$^{-1}$ (PR) and K (T2M).}
\label{tab:headline-results}
\small
\begin{tabular}{lllrrr}
\toprule
Mask & Variable & Metric & \modelshort{} & \methodname{} & Skill (\%) \\
\midrule
All & PR & CRPS $\downarrow$ & 1.591 & 1.079 & $+32.2$ \\
All & PR & RMSE $\downarrow$ & 3.448 & 2.565 & $+25.6$ \\
All & T2M & CRPS $\downarrow$ & 1.193 & 0.788 & $+34.0$ \\
All & T2M & RMSE $\downarrow$ & 2.239 & 1.738 & $+22.4$ \\
Land & PR & CRPS $\downarrow$ & 1.302 & 0.881 & $+32.3$ \\
Land & PR & RMSE $\downarrow$ & 3.083 & 2.294 & $+25.6$ \\
Land & T2M & CRPS $\downarrow$ & 1.936 & 1.303 & $+32.7$ \\
Land & T2M & RMSE $\downarrow$ & 3.212 & 2.584 & $+19.5$ \\
\bottomrule
\end{tabular}
\end{table}

\begin{figure}[!tbp]
\centering
\includegraphics[width=0.9\linewidth]{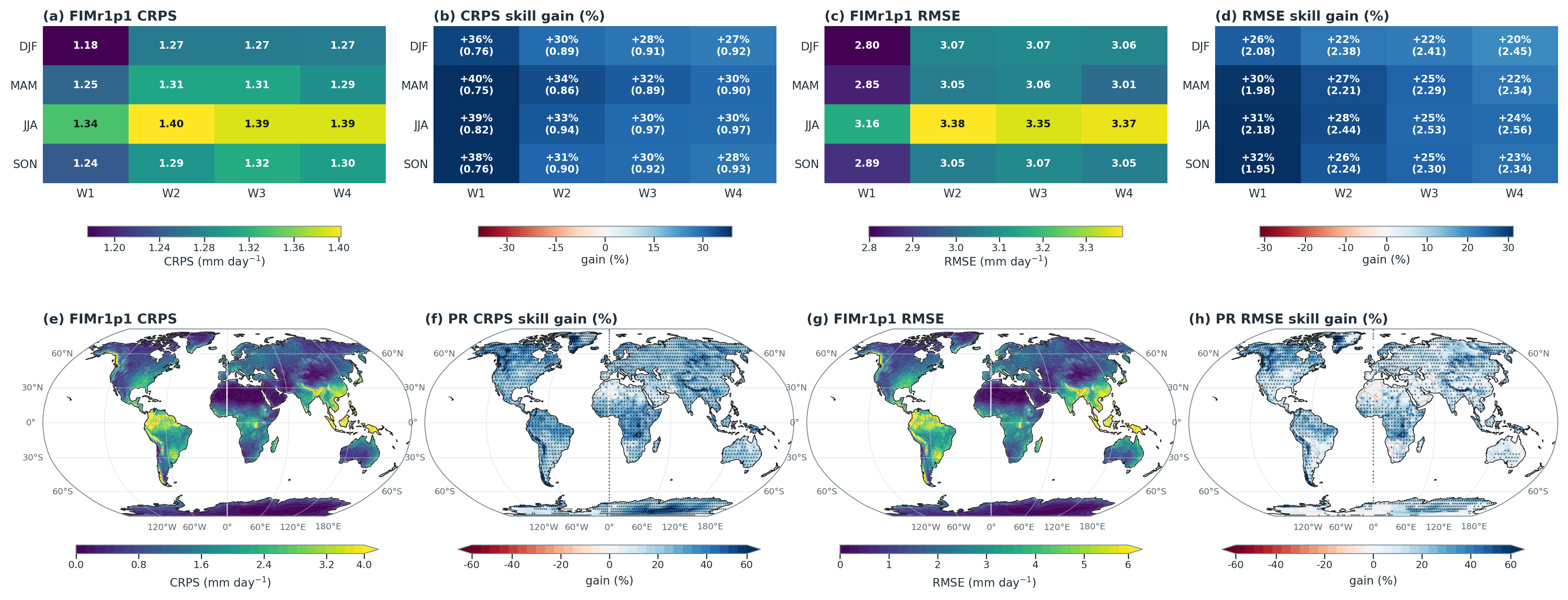}
\caption{Precipitation all-case skill versus \modelshort{}, 2021--2025. (a,~c) Season--lead matrices of raw \modelshort{} CRPS and RMSE; (b,~d) corresponding \methodname{} skill (\%, Eq.~\ref{eq:skill}) with the absolute \methodname{} score in parentheses; (e,~g) spatial maps of raw \modelshort{} CRPS and RMSE aggregated over seasons and leads; (f,~h) spatial maps of \methodname{} CRPS and RMSE skill. Blue denotes improvement over the dynamical baseline, red degradation; stippling marks gridpoints where the improvement has consistent sign in at least 95\% of the 16 season--lead cells; this is a descriptive sign-consistency summary, not a statistical significance test.}
\label{fig:pr-skill}
\end{figure}

\begin{figure}[!tbp]
\centering
\includegraphics[width=0.9\linewidth]{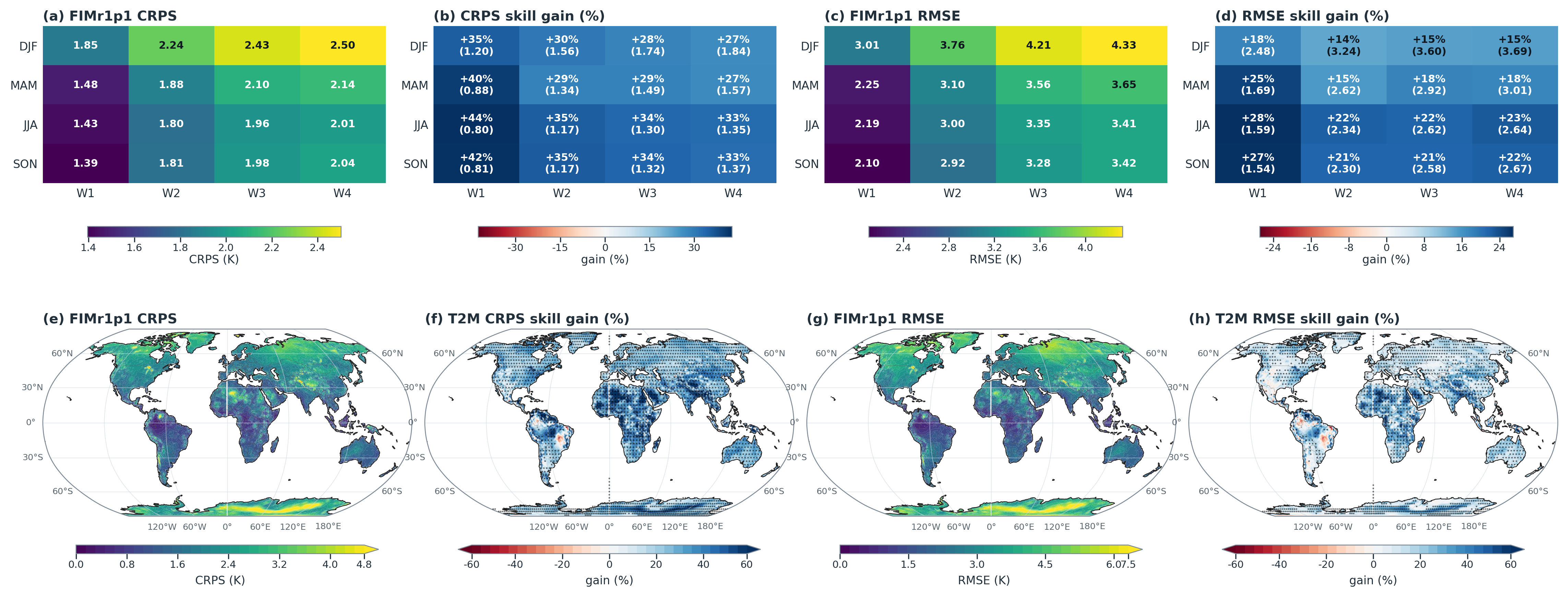}
\caption{2\,m temperature all-case skill versus \modelshort{}, 2021--2025, in the same layout as Fig.~\ref{fig:pr-skill}: season--lead matrices of raw \modelshort{} CRPS and RMSE with the corresponding \methodname{} skill (\%, Eq.~\ref{eq:skill}), and spatial maps of the same four quantities. Blue denotes improvement over the dynamical baseline, red degradation; stippling marks gridpoints where the improvement has consistent sign in at least 95\% of the 16 season--lead cells; this is a descriptive sign-consistency summary, not a statistical significance test.}
\label{fig:t2m-skill}
\end{figure}

Figures~\ref{fig:pr-skill} and \ref{fig:t2m-skill} resolve the production-configuration aggregates by season, lead, and geography. Every one of the 16 season--lead cells is improved for both variables and both masks: all-grid precipitation CRPS skill ranges from 27.7\% to 40.3\% (land 27.3--39.7\%), and all-grid temperature CRPS skill from 23.7\% to 48.1\% (land 26.6--43.8\%). The spatial maps show that gains are geographically coherent rather than driven by a few regions: precipitation improvements are largest where raw \modelshort{} CRPS is largest (tropical convergence zones, monsoon regions, orographic coasts), and temperature improvements are broad over land with the strongest corrections in regions of known systematic bias.

Table~\ref{tab:lead-mask-results} shows the lead dependence, again in the production configuration. Skill is largest at week~1, where the correction problem is most deterministic, but remains large and stable through weeks 2--4 (all-grid precipitation CRPS skill 38.1\%, 31.7\%, 30.2\%, 29.5\%), indicating that the system is not merely a short-lead bias corrector: the week-3/4 gains are the operationally interesting result. One qualification applies: the training loss up-weights weeks 3--4 (Sect.~3.2), so part of the long-lead advantage could in principle reflect capacity allocation rather than intrinsic capability, and this weighting has not been ablated. Week-1 skill nevertheless remains the largest of all leads, so the weighting did not simply trade short-lead skill for long-lead skill.

\begin{table}[!tbp]
\centering
\caption{Lead-dependent all-case skill aggregated over the four verifying seasons, under the same production configuration as Table~\ref{tab:headline-results}. CRPS and RMSE entries are percent score reductions relative to \modelshort{} (Eq.~\ref{eq:skill}; both scores are negatively oriented, so positive entries denote improved skill), and the values in parentheses give the raw four-member \modelshort{} score for that lead (mm\,day$^{-1}$ for PR, K for T2M), so the table can be used as an absolute reference.}
\label{tab:lead-mask-results}
\small
\begin{tabular}{lllcccc}
\toprule
Mask & Variable & Quantity & Week 1 & Week 2 & Week 3 & Week 4 \\
\midrule
All & PR & CRPS skill (\%) & 38.1 (1.45) & 31.7 (1.59) & 30.2 (1.64) & 29.5 (1.63) \\
All & PR & RMSE skill (\%) & 30.2 (3.12) & 25.4 (3.48) & 24.1 (3.57) & 23.2 (3.56) \\
All & T2M & CRPS skill (\%) & 41.5 (1.01) & 34.8 (1.23) & 34.0 (1.30) & 27.7 (1.34) \\
All & T2M & RMSE skill (\%) & 25.5 (1.76) & 22.0 (2.34) & 22.4 (2.53) & 20.8 (2.61) \\
Land & PR & CRPS skill (\%) & 38.1 (1.22) & 32.1 (1.30) & 30.2 (1.32) & 29.0 (1.26) \\
Land & PR & RMSE skill (\%) & 29.9 (2.87) & 26.1 (3.08) & 24.2 (3.14) & 22.5 (3.02) \\
Land & T2M & CRPS skill (\%) & 40.1 (1.57) & 32.3 (2.00) & 31.0 (2.12) & 29.6 (2.20) \\
Land & T2M & RMSE skill (\%) & 23.8 (2.51) & 18.0 (3.39) & 18.7 (3.66) & 19.0 (3.78) \\
\bottomrule
\end{tabular}
\end{table}

\subsection{Ablation of the physically informed stochastic prior}
\label{sec:results-noise}

\begin{figure}[!tbp]
\centering
\includegraphics[width=0.9\linewidth]{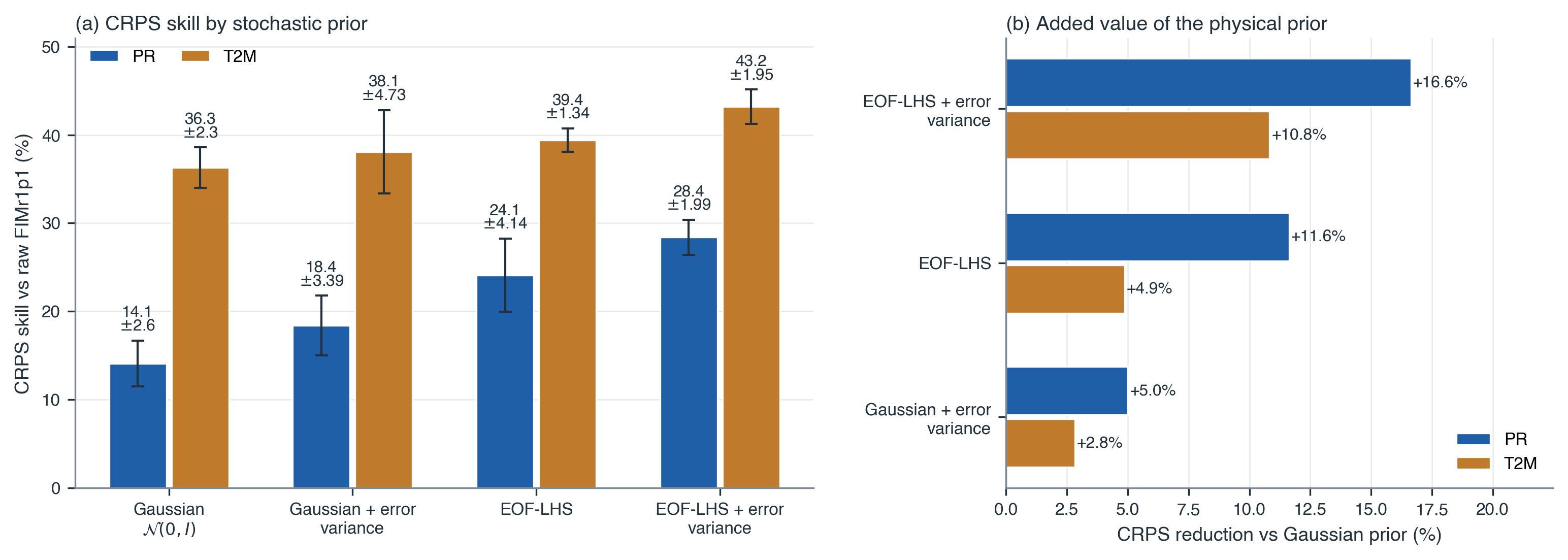}
\caption{Two-factor ablation of the stochastic-perturbation pipeline at fixed architecture, conditioning, ensemble size, and ODE solver, evaluated over 2021--2025 with 90 members. All arms use the same raw four-member \modelshort{} reference. (a) All-grid empirical-CRPS skill for the four configurations crossing Gaussian versus EOF-LHS prior with variance pathway off versus on. Each arm is trained and sampled under its own configuration and replicated with five checkpoints; bars show seed means and whiskers from a paired bootstrap with 100 resamples (approximate tail bounds). (b) Relative CRPS reduction with respect to the Gaussian/no-variance baseline.}
\label{fig:noise-ablation}
\end{figure}

The headline scores evaluate the full production system but do not identify how much skill comes from the stochastic prior itself. Figure~\ref{fig:noise-ablation} therefore reports the controlled ablation of Sect.~\ref{sec:design}: architecture, conditioning fields, training data and schedule, evaluation data (all weekly initializations of 2021--2025), ensemble size, and ODE steps are fixed, and only the stochastic-perturbation configuration changes across the four arms of the two-factor design (prior structure $\times$ error-variance pathway), with each arm trained and sampled under its own configuration. The production prior configuration was fixed from 2020 development scores before any evaluation-period data were examined; the ablation is a post-hoc diagnostic of that frozen choice, not a selection step.

All four arms are in-distribution, each trained and sampled under its own configuration, so no train--inference prior mismatch enters any comparison, and the two-factor design separates the two stochastic mechanisms: comparing the two priors at a fixed variance setting isolates the prior's covariance structure with spread control held constant, and comparing variance settings at a fixed prior isolates the learned error-variance pathway. Training variability is addressed directly: each arm is replicated with five independently trained checkpoints (20 in total), and all ablation scores are reported as seed means with paired-bootstrap intervals. The present ablation whiskers were estimated with 100 bootstrap replicates and should be read as approximate; stable tail-percentile bounds require at least 1{,}000 replicates. Within the prior factor, spatial EOF structure, regime conditioning, and the LHS design still enter together, so the structured prior is attributed as a unit; an unconditional-EOF arm is the natural next split.

The Gaussian baseline without the variance pathway already improves over raw \modelshort{} (precipitation CRPS skill $14.1\pm2.6\%$, temperature $36.3\pm2.3\%$), confirming that the conditional flow and the dynamical conditioning carry a large share of the skill on their own. The two mechanisms then add distinct increments. Activating the error-variance pathway under the Gaussian prior raises skill to $18.4\pm3.4\%$ and $38.1\pm4.7\%$ (relative CRPS reductions of 5.0\% and 2.8\% versus the baseline); switching to the EOF-LHS prior with the variance pathway off reaches $24.1\pm4.1\%$ and $39.4\pm1.3\%$ (11.6\% and 4.9\%); and the production combination reaches $28.4\pm2.0\%$ and $43.2\pm2.0\%$ (16.6\% and 10.8\%). For precipitation the two contributions are almost exactly additive ($5.0\%+11.6\%\approx16.6\%$), whereas for temperature the combination exceeds the sum of its parts ($2.8\%+4.9\%$ versus $10.8\%$), indicating a positive interaction between the structured initialization and the learned spread field. The structured prior is the larger of the two mechanisms for both variables, and the full-pipeline gaps over the Gaussian baseline (14.3 precipitation and 6.9 temperature skill points) are several times the between-seed confidence intervals, so optimization variability cannot account for the differences. We interpret the prior factor as the value of physically informed initialization: the EOF-LHS prior starts each member on a low-dimensional manifold of historically observed, regime-consistent anomaly structures that the velocity field learns to refine, and the larger benefit for precipitation is consistent with its greater sensitivity to displacement and organization uncertainty.

The full-pipeline ablation mean (28.4\% PR, 43.2\% T2M) is not numerically identical to the single-checkpoint production result (32.2\%, 34.0\%). The production table evaluates one checkpoint selected on 2020, whereas Fig.~\ref{fig:noise-ablation} averages five separately trained checkpoints per arm; the production checkpoint is not the ablation mean. The opposite-signed differences for PR and T2M, including the 9.2-point T2M gap, show that training-seed/checkpoint variation is scientifically material. We therefore use the ablation to compare factors within its paired multi-seed design and do not use it as a numerical reproduction of Table~\ref{tab:headline-results}; total uncertainty on the headline should include training-seed variability, which the single-checkpoint table does not.

The temperature response deserves separate comment, because the structured component supplies only $\rho^2\approx0.6\%$ of the initial T2M variance (Sect.~3.6), yet the prior factor adds 3.1 skill points at fixed spread control. A cross-channel pathway is one possible explanation: the two target channels are integrated jointly, and the precipitation channel carries stronger structured covariance. This mechanism is inferred, not demonstrated; a cross-channel ablation is required to establish it. The present supplement reports aggregate seed intervals but not the individual paired seed scores.

\subsection{Ensemble size and skill convergence}
\label{sec:results-ensemble}

\begin{figure}[!tbp]
\centering
\includegraphics[width=0.9\linewidth]{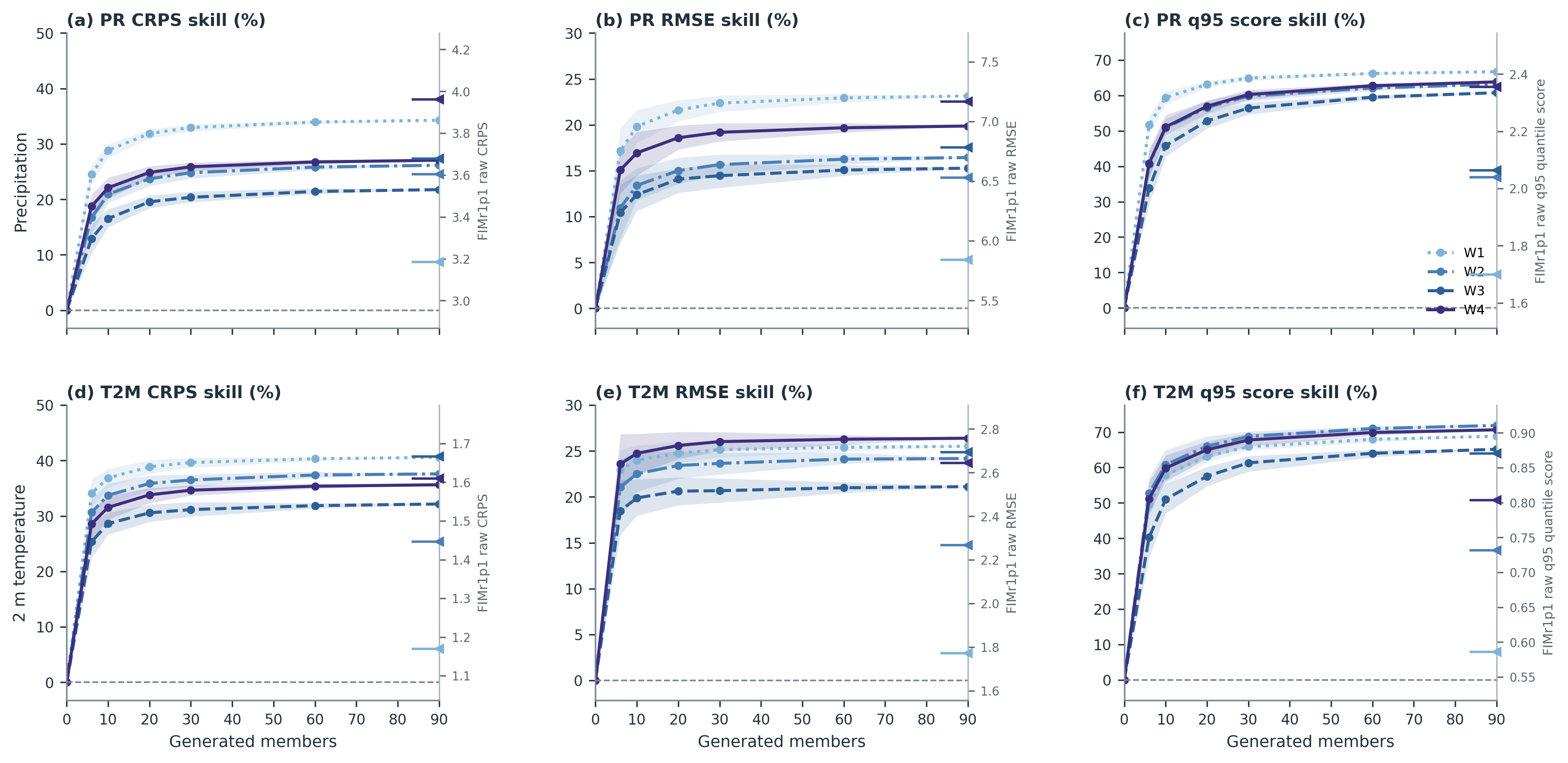}
\caption{Ensemble-size skill convergence on anomaly-selected extreme cases: CRPS, ensemble-mean RMSE, and type-7 $q_{0.95}$ QS skill for precipitation and 2\,m temperature. Every point uses the same raw four-member \modelshort{} reference; shading spans the 5th--95th percentile over repeated member subsets. The four-member QS point is maximum-dominated and is included only as a sampling diagnostic. The 90-member point combines forecast-distribution skill with a member-count effect. No exceedance mask is used in QS.}
\label{fig:member-convergence}
\end{figure}

A stochastic generator must also be characterized as an ensemble system, and for high-impact applications the operative question is specific: how many generated members are needed before the forecast distribution samples its extreme-event tail well enough to beat the dynamical baseline? Figure~\ref{fig:member-convergence} therefore evaluates CRPS, ensemble-mean RMSE, and 95th-percentile QS (Eq.~\ref{eq:pinball}) skill as functions of the number of generated members on the anomaly-selected extreme-event cases for all four lead weeks, including the longer leads where dynamical extreme-event guidance is weakest; CRPS evaluates the full ensemble distribution, RMSE the ensemble mean, and QS evaluates the ensemble's own upper-tail forecast, and none of the three uses the $q_{0.95}$ exceedance masks. Shading spans the 5th--95th percentile over repeated random member subsets.

Four features are relevant for practice. First, skill climbs steeply from the smallest subsets. Second, CRPS and RMSE skill largely saturate by roughly 20--30 members, with the 90-member configuration inside the plateau. Third, long-lead CRPS remains positive after convergence, although probability reliability is a separate question. Fourth, $q_{0.95}$ QS continues to improve beyond the mean-score plateau and reaches roughly 60--70\% skill at 90 members. The four-member QS values (34--53\% skill) are shown only to trace the curve: under the stated linear convention they are dominated by the ensemble maximum and should not be interpreted as well-resolved 95th-percentile forecasts. The adequately sampled 90-member tail result, which necessarily includes a member-count contribution, is the interpretable upper-tail evidence.

Mean skill is, however, the wrong yardstick for sizing a probabilistic ensemble, and its early saturation should not be read as ``30 members suffice.'' The same saturation appears in the all-case mean-skill curves (supplement): ensemble-mean RMSE and CRPS measure the bulk of the forecast distribution, whose member-sampling error shrinks rapidly with ensemble size $M$. Tail probabilities obey different statistics. An $M$-member ensemble can only issue exceedance probabilities quantized in steps of $1/M$, which is 25\% for the four-member dynamical ensemble. The sampling standard error of an estimated exceedance probability $p$ is $\sqrt{p(1-p)/M}$, which for a 5\% tail event and $M=8$ is $\pm7.7\%$, larger than the probability being estimated; resolving a local $q_{0.99}$ event at all requires $O(100)$ members. Ensemble size therefore continues to pay off precisely where the mean-skill curves of Fig.~\ref{fig:member-convergence} have flattened: in the sharpness and stability of rare-event probabilities and extreme quantiles, the quantities that carry forecast value for high-impact decisions \citep{Richardson2001}.

Additional generated members require repeated learned-flow integrations rather than coupled dynamical integrations, so enlargement is expected to be much cheaper after training. No standardized hardware benchmark is reported here, however, and we therefore make no numerical speedup claim. The evidence supports stable mean CRPS/RMSE estimates by roughly 20--30 members and improved $q_{0.95}$ diagnostics at 90 members; it does not demonstrate reliable $q_{0.99}$ or compound-event probabilities, nor statistical independence of additional members beyond the sampled prior modes.

\begin{figure}[!tbp]
\centering
\includegraphics[width=0.9\linewidth]{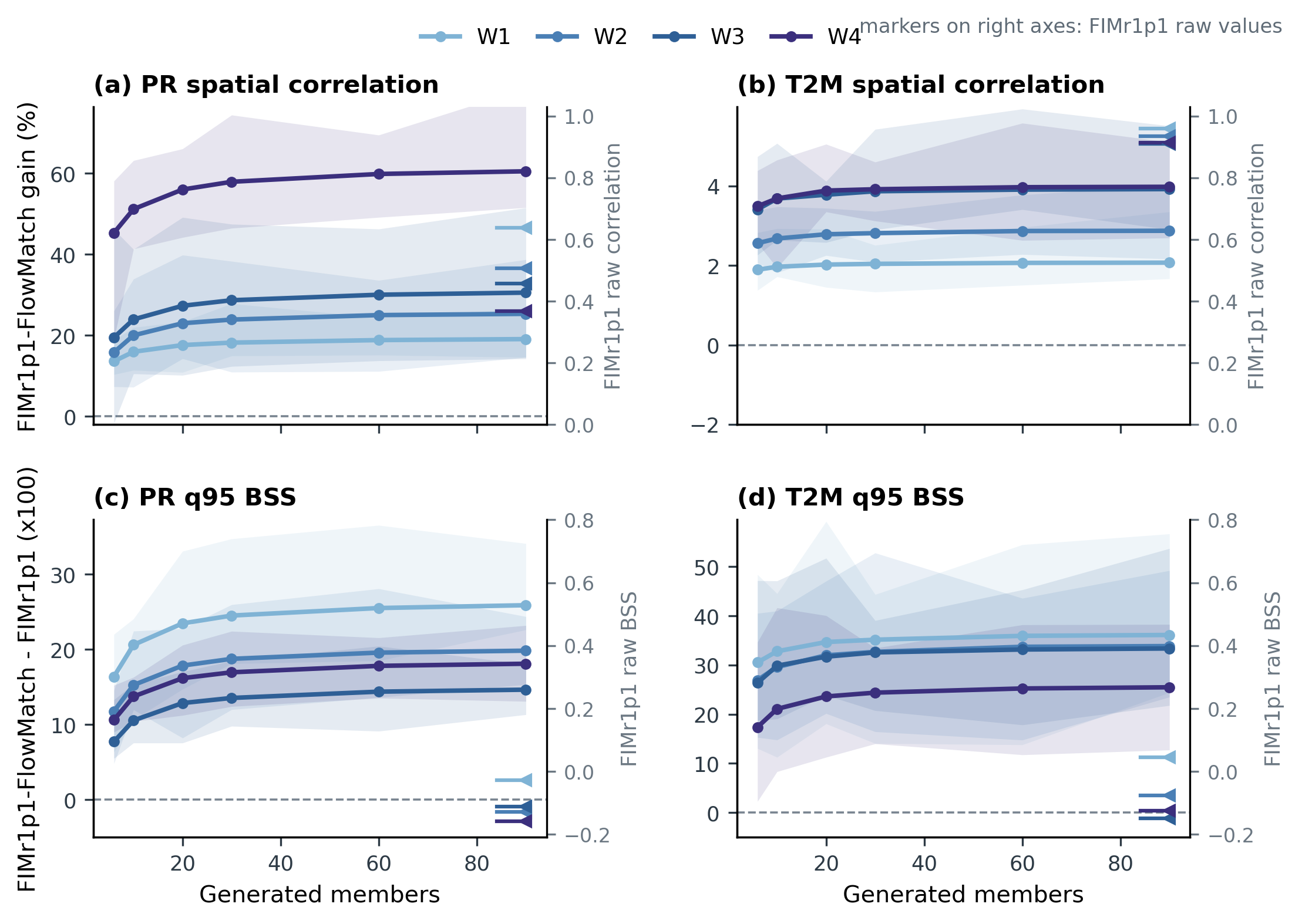}
\caption{Member-count convergence of pattern and event-probability skill on the same anomaly-selected extreme cases as Fig.~\ref{fig:member-convergence}, scored against the same raw four-member \modelshort{} reference, so the four-member point corresponds to the four-versus-four protocol of Table~\ref{tab:extreme-event-results}. (a,~b) Relative gain in extreme-case spatial correlation of the ensemble mean for precipitation and 2\,m temperature; (c,~d) gain in local $q_{0.95}$ exceedance BSS ($\times$100). Curves show lead weeks 1--4, shading spans the 5th--95th percentile over resampling, the dashed line marks zero gain, and dotted lines (right axes) show the raw \modelshort{} values of the underlying quantity. Precipitation correlation gains are largest at week 4, where raw dynamical pattern skill is lowest, whereas temperature gains are small because raw correlations are already 0.91--0.96.}
\label{fig:member-corr-bss}
\end{figure}

Figure~\ref{fig:member-corr-bss} extends the member sweep to spatial correlation and unconditional exceedance BSS. Precipitation pattern correlation improves most at week~4, where the raw value is lowest (0.37 to 0.54 at four members); temperature gains are smaller because raw correlations are already 0.91--0.96. The probability result is more qualified. Temperature BSS is positive at every lead after refinement, but precipitation is positive at week~1 and approximately neutral at weeks 2--4. Thus the large conditional extreme-case CRPS/QS improvements do not demonstrate calibrated operational precipitation probabilities at long lead. Reliability diagrams and Brier reliability--resolution--uncertainty decompositions are needed before such a claim can be made.

\subsection{Skill on observed extremes}
\label{sec:results-extremes}

\begin{figure}[!tbp]
\centering
\includegraphics[width=0.9\linewidth]{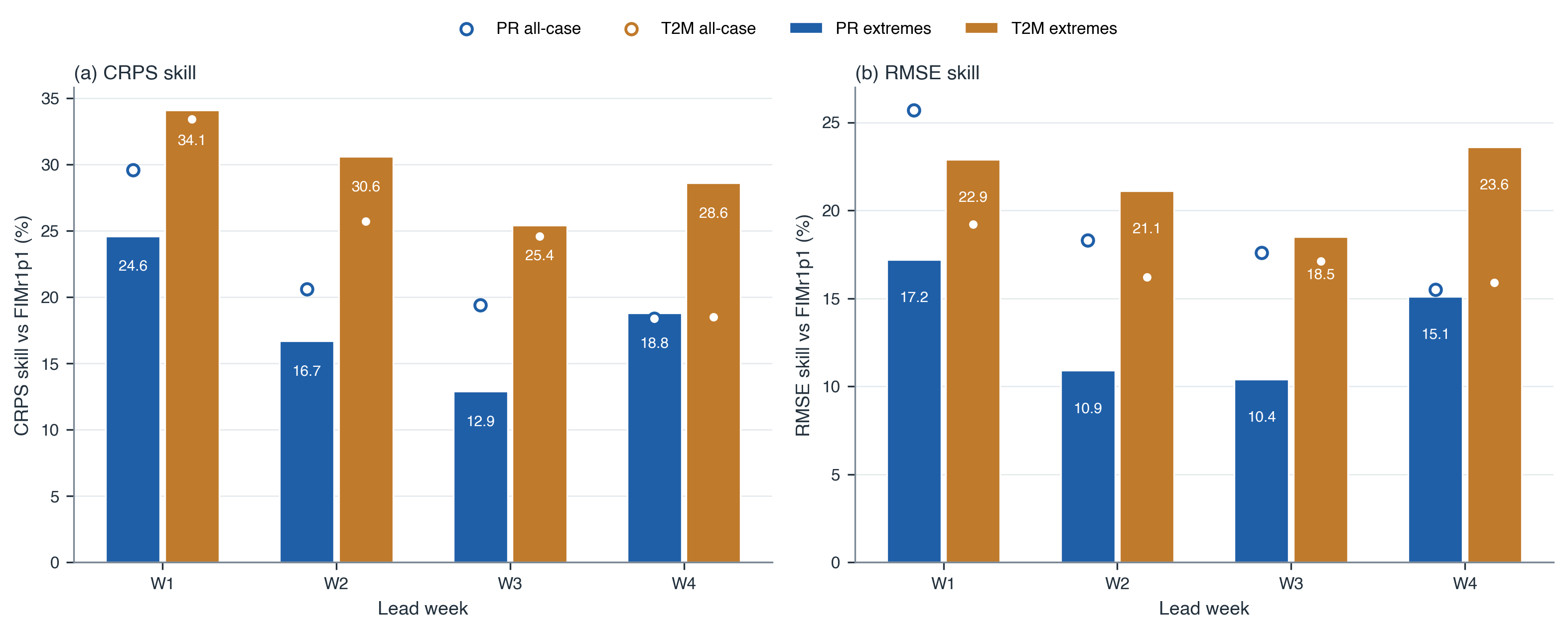}
\caption{Conditional CRPS and RMSE skill on 120 observed-extreme cases under the four-versus-four size-matched protocol. Open markers show corresponding all-case empirical-score skill. Cases are selected by observed anomaly magnitude; no $q_{0.95}$ mask is used.}
\label{fig:extreme-skill}
\end{figure}

Figure~\ref{fig:extreme-skill} and Table~\ref{tab:extreme-event-results} evaluate 120 observed-extreme cases (60 precipitation and 60 2\,m temperature) from 2021--2025. This is explicitly \emph{conditional verification}: it asks how the forecasts behaved given that an extreme occurred, not whether the system identifies rare events over all issue dates. The same-initialization four-versus-four size-matched comparison yields positive empirical-CRPS and RMSE skill at every lead. For precipitation, mean CRPS skill ranges from 12.9\% to 24.6\% and mean RMSE skill from 10.4\% to 17.2\%; for temperature, the ranges are 25.4--34.1\% and 18.5--23.6\%. Four-member $q_{0.95}$ QS skill is 33.8--51.6\% for precipitation and 40.2--52.7\% for temperature, but these values are maximum-dominated under the stated quantile convention and are secondary to CRPS. Spread/RMSE ratios of 0.51--0.65 indicate systematic underdispersion, and the unconditional BSS result is much weaker for precipitation after week~1. The evidence therefore supports conditional case performance and improved tail representation, not calibrated rare-event probabilities. Rank histograms in the supplement document the dispersion deficit; full reliability diagrams and Brier decomposition remain required for an operational probability claim.

\begin{table}[!tbp]
\centering
\caption{Conditional skill given 120 observed-extreme cases (60 precipitation, 60 2\,m temperature), 2021--2025. Both systems use four members. Values use standard empirical CRPS; QS$_{0.95}$ uses type-7 interpolation and is maximum-dominated at $M=4$. Skill entries are percentages with 5th--95th percentile (90\%) paired case-resampling intervals. These scores do not measure unconditional rare-event detection or probability reliability.}
\label{tab:extreme-event-results}
\small
\resizebox{\textwidth}{!}{%
\begin{tabular}{llcccccc}
\toprule
Variable & Lead & \modelshort{} CRPS & CRPS skill (\%) & \modelshort{} RMSE & RMSE skill (\%) & \modelshort{} QS$_{0.95}$ & QS$_{0.95}$ skill (\%) \\
\midrule
PR & Week 1 & 3.19 & 24.6 (22.8--26.2) & 5.84 & 17.2 (15.0--19.6) & 1.70 & 51.6 (49.1--54.2) \\
PR & Week 2 & 3.60 & 16.7 (13.7--19.3) & 6.53 & 10.9 (7.6--14.0) & 2.04 & 40.5 (36.7--44.4) \\
PR & Week 3 & 3.68 & 12.9 (10.2--15.5) & 6.78 & 10.4 (7.1--13.5) & 2.06 & 33.8 (29.2--38.1) \\
PR & Week 4 & 3.96 & 18.8 (16.1--20.9) & 7.17 & 15.1 (12.5--17.4) & 2.36 & 40.8 (36.8--44.7) \\
T2M & Week 1 & 1.17 & 34.1 (31.5--36.6) & 1.77 & 22.9 (20.7--25.1) & 0.59 & 49.6 (45.4--54.0) \\
T2M & Week 2 & 1.45 & 30.6 (28.1--33.2) & 2.27 & 21.1 (18.0--24.1) & 0.73 & 52.7 (46.5--57.1) \\
T2M & Week 3 & 1.67 & 25.4 (22.4--28.8) & 2.69 & 18.5 (15.9--21.8) & 0.87 & 40.2 (34.5--45.7) \\
T2M & Week 4 & 1.61 & 28.6 (25.7--31.3) & 2.65 & 23.6 (20.8--26.8) & 0.80 & 51.2 (46.4--55.9) \\
\bottomrule
\end{tabular}}
\end{table}

\begin{figure}[!tbp]
\centering
\includegraphics[width=1\linewidth]{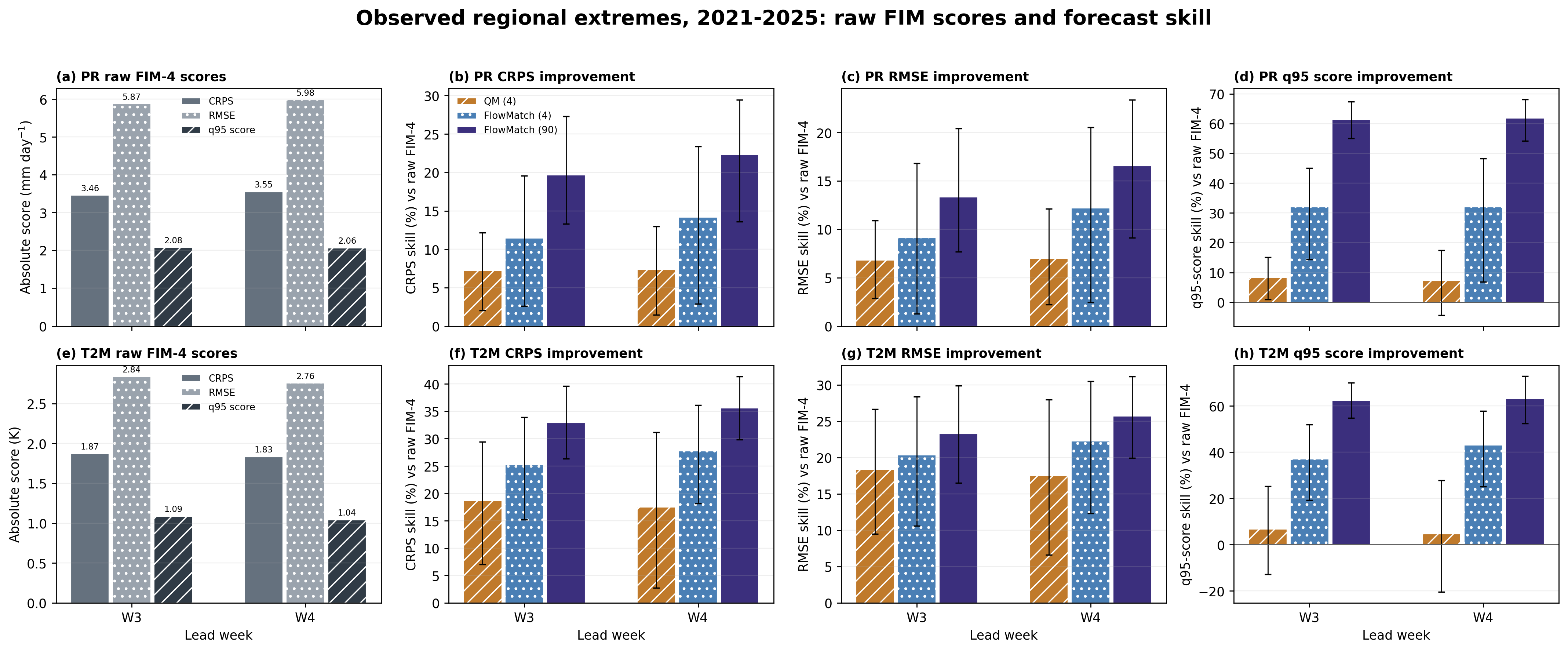}
\caption{Conditional observed-extreme cases at weeks 3--4: frozen 1999--2019 quantile mapping versus generative refinement. Error bars are paired case-resampling intervals. The four-member type-7 $q_{0.95}$ values are maximum-dominated; FlowMatch-90 provides an adequately sampled tail but includes a member-count advantage. This figure is not an all-case postprocessing comparison.}
\label{fig:extreme-qm}
\end{figure}

Figure~\ref{fig:extreme-qm} provides a focused conventional-baseline check on the observed regional extreme cases at weeks 3 and 4. QM recovers part of the mean-error improvement (CRPS gains of about 7\% for precipitation and 18--19\% for temperature), while \methodname{} has larger CRPS and RMSE gains in these cases. Four-member QS differences are reported but are not treated as decisive because both QM-4 and FlowMatch-4 have under-resolved 95th percentiles. At 90 generated members, $q_{0.95}$ QS skill reaches 61--63\%, but this comparison includes the ensemble-size contribution. Because QM is evaluated only on this conditional extreme subset, these results do not establish all-case superiority to standard postprocessing; an all-case QM or EMOS comparison is still needed for that broader claim.

\subsection{High-impact event case studies}
\label{sec:results-events}

Beyond the systematic threshold subset, we examine named high-impact cases drawn from a pre-defined catalog assembled from evaluation-period disasters documented by official agencies and major-outlet reporting. The two main-text cases were chosen for distinct variables, regions, and drivers: the July 2022 UK heatwave \citep{MetOffice2022Heatwave} and the December 2022--January 2023 California atmospheric-river sequence \citep{CW3E2023CaliforniaAR}. A shared Pakistan miss is shown in the supplement. Because the complete case-level catalog and inclusion/exclusion audit are not included in the present supplement, these maps are illustrative rather than representative evidence. They use the initialization whose week-4 target window overlaps the documented event window. For a single event the maps have no replication dimension, and stippling marks fixed gain thresholds ($>$10\% CRPS skill; $>+5$ BSS gain $\times$100), not statistical significance.

\begin{figure}[!tbp]
\centering
\includegraphics[width=0.9\linewidth]{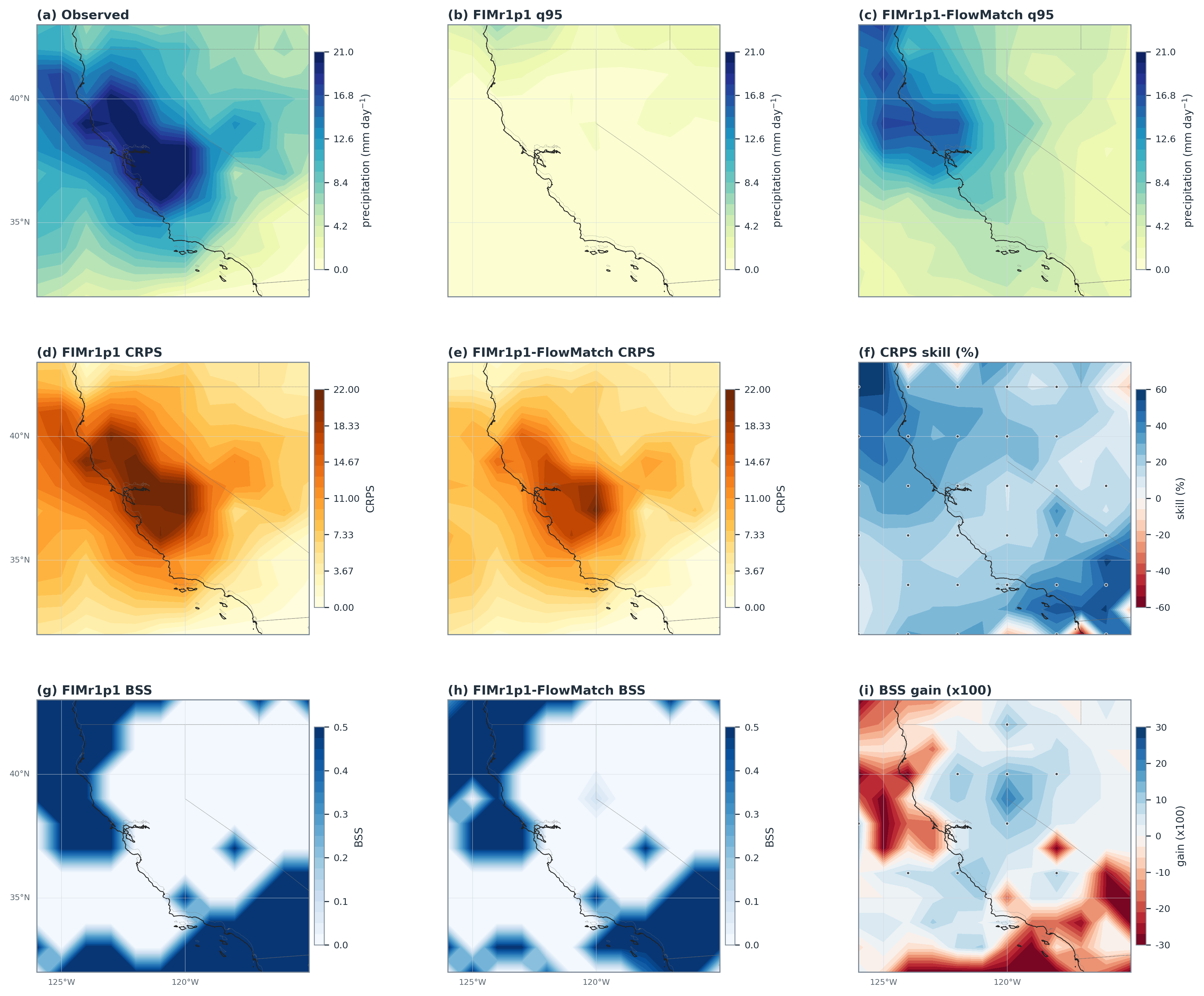}
\caption{California atmospheric-river precipitation case study (January 2023 storm week, week-4 lead). (a) Observed weekly precipitation; (b,~c) \modelshort{} and \methodname{} ensemble 95th percentiles; (d,~e) event-week CRPS for each system and (f) the resulting CRPS skill (\%, blue = improvement, stippled where $>10\%$); (g,~h) heavy-rain exceedance BSS for each system and (i) the BSS gain ($\times100$, stippled where the gain exceeds $+5$).}
\label{fig:event-pr}
\end{figure}

For the January 2023 California atmospheric-river week (Fig.~\ref{fig:event-pr}), the \methodname{} 95th-percentile precipitation field concentrates heavy-rain risk along the coastal and Sierra Nevada orography where the observed maxima occurred, whereas the raw \modelshort{} q95 is weaker and more diffuse. Event-area CRPS is reduced over much of the affected region, and BSS for local heavy-rain exceedance improves over parts of the coastal mountains; stippling marks a fixed gain threshold, not statistical robustness.

\begin{figure}[!tbp]
\centering
\includegraphics[width=0.9\linewidth]{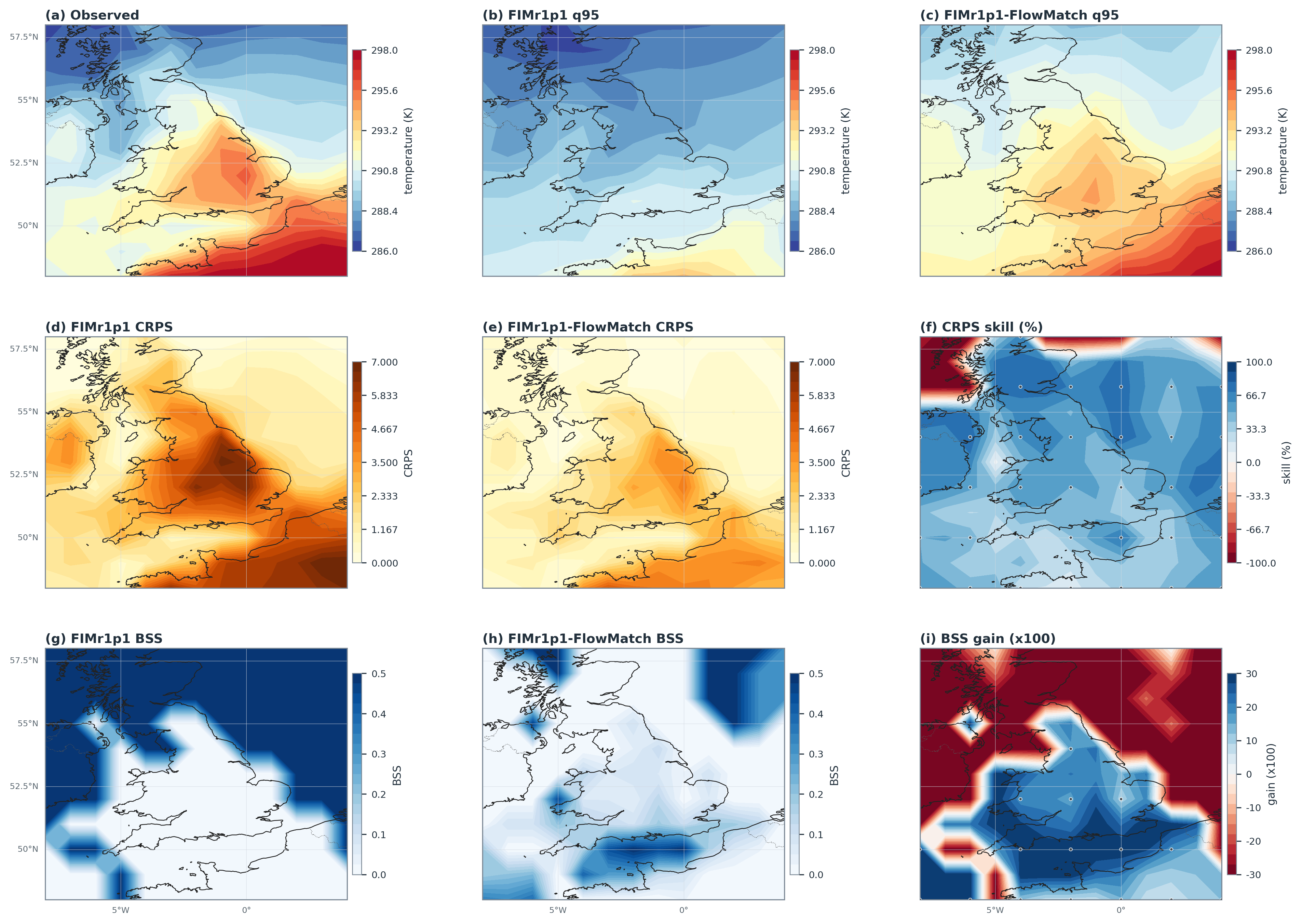}
\caption{UK heatwave 2\,m temperature case study (July 2022 event week, week-4 lead), in the same layout as Fig.~\ref{fig:event-pr}: observed T2M, \modelshort{} and \methodname{} ensemble 95th percentiles, event-week CRPS for each system with the resulting CRPS skill, and hot-event exceedance BSS for each system with the BSS gain.}
\label{fig:event-t2m}
\end{figure}

For the July 2022 UK heatwave (Fig.~\ref{fig:event-t2m}), raw \modelshort{} q95 temperatures remain several kelvin too cool over England at week-4 lead. The refined ensemble shifts the hot tail toward the observed anomaly pattern and reduces event-area CRPS over much of the region. These descriptive maps show plausible event-level behavior but do not establish reliability or representative performance across all extremes.

Not every high-impact event is captured. For the August 2022 Pakistan monsoon floods \citep{Otto2023PakistanFloods}, neither the raw \modelshort{} ensemble nor \methodname{} assigned elevated heavy-rain probabilities over the Indus basin at weeks 3--4. In this case both systems missed the observed high-impact precipitation event; the raw dynamical ensemble placed its heavier-rain signal southeast of the observed rainfall maximum, and the refined ensemble did not correct that event-scale displacement. The supplementary Pakistan maps (Fig.~S4) document the shared miss in the same layout as Figs.~\ref{fig:event-pr}--\ref{fig:event-t2m}. We view reporting such cases as essential for a fair assessment of ML-based refinement, but they should be read as evidence that some events remain difficult for both systems and need further diagnosis, not as a general limit on learned correction of weak or displaced dynamical signals.

\section{Discussion}
\label{sec:discussion}

The main scientific value of this framework is its hybrid interpretation. \modelshort{} supplies a dynamically consistent prior distribution, while the neural model learns a conditional transformation toward observations. This division of labor matches the complementary weaknesses identified in the introduction: the dynamical core contributes the physically consistent evolution of the slow components that a few decades of reanalysis cannot teach a network, while the learned refinement supplies what the dynamical system cannot afford, bias-aware distributional correction and dense, inexpensive ensemble sampling. This differs both from a purely data-driven rollout and from scalar bias correction: the network sees spatially complete \modelshort{} forecast summaries, recent observed environmental predictors, and static geography, and produces a new ensemble distribution whose spatial covariance is inherited jointly from the learned flow and from the physically informed stochastic prior.

Several design choices plausibly contribute to the observed behavior. The \modelshort{}-residual target for temperature focuses learning on forecast error, while direct precipitation prediction avoids a signed residual for an intermittent, displaced field. Static geography and global-context attention provide persistent and large-scale context. The five-seed factorial ablation supports a contribution from both the structured prior and variance pathway, although regime conditioning and LHS remain bundled and the cross-channel interpretation is untested. The conditional week-3/4 QM comparison shows added value beyond a marginal correction for those cases, but it does not replace an all-case postprocessing benchmark. Member-count experiments show that the generated distribution can be sampled more densely after training; without a standardized runtime benchmark or calibration analysis, we interpret this as a computational design opportunity rather than an operational probability product.

Equally important is what remains unresolved. Refinement is conditional on its inputs, but the case studies show both sides of that statement. When the raw ensemble contains a poorly calibrated signal, such as a too-cool heat dome or an over-smoothed atmospheric-river corridor, the learned transformation can relocate probability mass, as in the UK and California cases. The August 2022 Pakistan floods are different: at weeks 3--4, both the raw \modelshort{} ensemble and \methodname{} missed the observed Indus-basin heavy rainfall. That shared miss identifies a class of high-impact failures requiring further study. Because reliability has not been established and the system is underdispersive, the operationally defensible conclusion is improved probabilistic scores and tail representation for many cases, not reliable tail-risk probabilities.

The extreme-event results must be interpreted with the logic of probability forecasting. At subseasonal leads a well-calibrated ensemble mean should be smoother than individual outcomes; that is a feature, not a failure. The correct verification question for rare events is whether the forecast distribution assigns elevated, spatially relevant probability to the event and whether the observed value falls within a plausible forecast tail. On both the 120-case extreme-event catalog and the named case studies, \methodname{} improves on the raw dynamical ensemble under exactly these distribution-based criteria, while the four-member event-case spread/RMSE ratios (0.51--0.64) and the full-ensemble spread/RMSE ratios of 0.80 (PR) and 0.70 (T2M) indicate that residual overconfidence, not signal absence, is the dominant remaining calibration error. Until explicit spread or probability calibration is applied, the system should therefore be read as delivering improved but underdispersive probabilistic forecasts, not calibrated tail risk.

\section{Limitations and Future Work}
\label{sec:limitations}

Ten limitations qualify these results. First, the generated ensemble is underdispersive, with spread/RMSE ratios of 0.51--0.65 in the size-matched configurations, so raw tail probabilities are overconfident; reliability diagrams, Brier decomposition, and explicit spread calibration remain necessary before decision use. Second, equal member count does not eliminate distribution-dependent finite-$M$ bias in empirical CRPS, and the 90-versus-four production score includes a member-count effect. Third, $q_{0.95}$ from four members is maximum-dominated and convention-sensitive; only the 90-member tail is adequately sampled, and its gain cannot be separated from ensemble size in the present results. Fourth, the structured prior and variance tempering are partly heuristic and the lead weights and flow-time sampler lack dedicated ablations. Fifth, the variance mechanism adjusts initial spread only rather than injecting stochasticity continuously along the trajectory. Sixth, verification truth is uncertain: GPCP PR has known limitations over ocean and complex terrain, while ERA5 is both a predictor source and T2M verification truth. Seventh, although the prior ablation uses five seeds per arm, the headline production table uses one checkpoint; the 9.2-point difference between its T2M skill and the multi-seed full-pipeline mean demonstrates that seed uncertainty matters. Eighth, member-level spatial and multivariate coherence has not been verified: current evidence is gridpoint scores, ensemble-mean patterns, and quantile maps, and the prior encodes no PR--T2M covariance. A variogram or energy score, spatial spectra, and cross-variable dependence diagnostic are required before spatial coherence is claimed. Ninth, the conventional baseline suite remains incomplete because QM is reported only on week-3/4 conditional extreme cases; an all-case QM/EMOS or probabilistic neural-postprocessor comparison is needed for broad architecture claims. Tenth, observed predictors and indices are retrospective products without a point-in-time publication-vintage audit. The 28-day RMM lag is a modeling choice, not proof of operational availability, and its physical sensitivity relative to a near-initialization RMM state is untested. The system is therefore a research hindcast postprocessor.

Natural next ablations are no-\modelshort{}-conditioning, no-static-geography, no-global-cross-attention, variance tempering on/off, learned versus fixed EOF mixture weights, and absolute versus residual temperature targets. Longer-term extensions include learning the stochastic prior directly, training with differentiable CRPS in physical units, conservation-aware precipitation constraints, SDE rather than ODE sampling, multi-model dynamical conditioning to reduce single-model signal misses, and extending the target set to circulation variables.

\section{Conclusion}
\label{sec:conclusion}

We presented \methodname{}, a dynamical-model-informed stochastic flow-matching framework for global probabilistic subseasonal forecasting of weekly precipitation and 2\,m temperature. On held-out 2021--2025 data, standard empirical CRPS improves by 22.0\% (precipitation) and 25.6\% (temperature) in a same-initialization four-versus-four comparison and by 32.2\% and 34.0\% in the 90-versus-four production comparison. The first controls direct member count but retains distribution-dependent finite-$M$ bias; the second includes a member-count contribution. Conditional CRPS and RMSE also improve across the 120 observed-extreme cases, while unconditional long-lead precipitation BSS is near neutral and the ensemble remains underdispersive. These results support improved probabilistic scores and tail representation, not calibrated operational rare-event probabilities or verified member-level spatial coherence. A documented Pakistan flood case marks a shared high-impact miss by both systems.

The evidence supports three narrower conclusions: machine learning can improve a physics-based subseasonal ensemble without replacing its dynamical guidance; within a paired five-seed ablation, the physically informed prior and learned variance pathway each add skill relative to Gaussian counterparts; and inexpensive generated members improve sampling of the forecast distribution. Broader claims about unbiased CRPS superiority, calibrated rare-event probabilities, operational information availability, or multivariate spatial coherence require the additional analyses identified above.

\section*{Code and data availability}

The evaluation code, the frozen score files underlying every table and figure, and the figure-generation scripts are provided to reviewers at submission; the full training and inference code will be released publicly upon publication. \modelshort{} forecasts are available from the SubX archive \citep{Pegion2019SubX}; GPCP, ERA5, OISST, Copernicus salinity, and C3S soil-moisture data are available from their respective providers.

\bibliographystyle{plainnat}
\bibliography{references}

@article{Cannon2015,
  author = {Cannon, Alex J. and Sobie, Stephen R. and Murdock, Trevor Q.},
  title = {Bias correction of {GCM} precipitation by quantile mapping: How well do methods preserve changes in quantiles and extremes?},
  journal = {Journal of Climate},
  year = {2015},
  volume = {28},
  pages = {6938--6959},
  doi = {10.1175/JCLI-D-14-00754.1}
}

@article{Maraun2016,
  author = {Maraun, Douglas},
  title = {Bias correcting climate change simulations - a critical review},
  journal = {Current Climate Change Reports},
  year = {2016},
  volume = {2},
  pages = {211--220},
  doi = {10.1007/s40641-016-0050-x}
}

@article{Koster2004,
  author = {Koster, Randal D. and Dirmeyer, Paul A. and Guo, Zhichang and Bonan, Gordon and Chan, Edmond and Cox, Peter and Gordon, C. T. and Kanae, Shinjiro and Kowalczyk, Eva and Lawrence, David and Liu, Ping and Lu, Cheng-Hsuan and Malyshev, Sergey and McAvaney, Bryant and Mitchell, Ken and Mocko, David and Oki, Taikan and Oleson, Keith and Pitman, Andrew and Sud, Y. C. and Taylor, Christopher M. and Verseghy, Diana and Vasic, Ratko and Xue, Yongkang and Yamada, Tomohito},
  title = {Regions of strong coupling between soil moisture and precipitation},
  journal = {Science},
  year = {2004},
  volume = {305},
  pages = {1138--1140},
  doi = {10.1126/science.1100217}
}

@article{ZhuNewell1998,
  author = {Zhu, Yong and Newell, Reginald E.},
  title = {A proposed algorithm for moisture fluxes from atmospheric rivers},
  journal = {Monthly Weather Review},
  year = {1998},
  volume = {126},
  pages = {725--735},
  doi = {10.1175/1520-0493(1998)126<0725:APAFMF>2.0.CO;2}
}

@article{Li2016SSS,
  author = {Li, Laifang and Schmitt, Raymond W. and Ummenhofer, Caroline C. and Karnauskas, Kristopher B.},
  title = {North {Atlantic} salinity as a predictor of {Sahel} rainfall},
  journal = {Science Advances},
  year = {2016},
  volume = {2},
  pages = {e1501588},
  doi = {10.1126/sciadv.1501588}
}

@article{VitartRobertson2018,
  author = {Vitart, Fr\'ed\'eric and Robertson, Andrew W.},
  title = {The sub-seasonal to seasonal prediction project ({S2S}) and the prediction of extreme events},
  journal = {npj Climate and Atmospheric Science},
  year = {2018},
  volume = {1},
  pages = {3},
  doi = {10.1038/s41612-018-0013-0}
}

@article{Bauer2015,
  author = {Bauer, Peter and Thorpe, Alan and Brunet, Gilbert},
  title = {The quiet revolution of numerical weather prediction},
  journal = {Nature},
  year = {2015},
  volume = {525},
  pages = {47--55},
  doi = {10.1038/nature14956}
}

@article{LeutbecherPalmer2008,
  author = {Leutbecher, Martin and Palmer, Tim N.},
  title = {Ensemble forecasting},
  journal = {Journal of Computational Physics},
  year = {2008},
  volume = {227},
  pages = {3515--3539},
  doi = {10.1016/j.jcp.2007.02.014}
}

@article{Stan2017,
  author = {Stan, Cristiana and Straus, David M. and Frederiksen, Jorgen S. and Lin, Hai and Maloney, Eric D. and Schumacher, Courtney},
  title = {Review of tropical-extratropical teleconnections on intraseasonal time scales},
  journal = {Reviews of Geophysics},
  year = {2017},
  volume = {55},
  pages = {902--937},
  doi = {10.1002/2016RG000538}
}

@article{Kochkov2024,
  author = {Kochkov, Dmitrii and Yuval, Janni and Langmore, Ian and Norgaard, Peter and Smith, Jamie and Mooers, Griffin and Kl{\"o}wer, Milan and Lottes, James and Rasp, Stephan and D{\"u}ben, Peter and Hatfield, Sam and Battaglia, Peter and Sanchez-Gonzalez, Alvaro and Willson, Matthew and Brenner, Michael P. and Hoyer, Stephan},
  title = {Neural general circulation models for weather and climate},
  journal = {Nature},
  year = {2024},
  volume = {632},
  pages = {1060--1066},
  doi = {10.1038/s41586-024-07744-y}
}

@article{Mouatadid2023,
  author = {Mouatadid, Soukayna and Orenstein, Paulo and Flaspohler, Genevieve and Cohen, Judah and Oprescu, Miruna and Fraenkel, Ernest and Mackey, Lester},
  title = {Adaptive bias correction for improved subseasonal forecasting},
  journal = {Nature Communications},
  year = {2023},
  volume = {14},
  pages = {3482},
  doi = {10.1038/s41467-023-38874-y}
}

@article{WatsonParris2021,
  author = {Watson-Parris, Duncan},
  title = {Machine learning for weather and climate are worlds apart},
  journal = {Philosophical Transactions of the Royal Society A},
  year = {2021},
  volume = {379},
  pages = {20200098},
  doi = {10.1098/rsta.2020.0098}
}

@article{Price2025,
  author = {Price, Ilan and Sanchez-Gonzalez, Alvaro and Alet, Ferran and Andersson, Tom R. and El-Kadi, Andrew and Masters, Dominic and Ewalds, Timo and Stott, Jacklynn and Mohamed, Shakir and Battaglia, Peter and Lam, Remi and Willson, Matthew},
  title = {Probabilistic weather forecasting with machine learning},
  journal = {Nature},
  year = {2025},
  volume = {637},
  pages = {84--90},
  doi = {10.1038/s41586-024-08252-9}
}

@misc{GPCPdaily2017,
  author = {Adler, Robert and Wang, Jian-Jian and Sapiano, Mathew and Huffman, George and Bolvin, David and Nelkin, Eric and {NOAA CDR Program}},
  title = {Global Precipitation Climatology Project ({GPCP}) Climate Data Record ({CDR}), Version 1.3 (Daily)},
  year = {2017},
  publisher = {NOAA National Centers for Environmental Information},
  doi = {10.7289/V5RX998Z},
  url = {https://www.ncei.noaa.gov/access/metadata/landing-page/bin/iso?id=gov.noaa.ncdc:C00999},
  note = {NCEI DSI 3631\_02; global daily fields for 1999--2025; accessed 2026-07-13}
}

@article{Huffman2001,
  author = {Huffman, George J. and Adler, Robert F. and Morrissey, Mark M. and Bolvin, David T. and Curtis, Scott and Joyce, Robert and McGavock, Brad and Susskind, Joel},
  title = {Global Precipitation at One-Degree Daily Resolution from Multisatellite Observations},
  journal = {Journal of Hydrometeorology},
  year = {2001},
  volume = {2},
  number = {1},
  pages = {36--50},
  doi = {10.1175/1525-7541(2001)002<0036:GPAODD>2.0.CO;2}
}

@article{BarnstonLivezey1987,
  author = {Barnston, Anthony G. and Livezey, Robert E.},
  title = {Classification, Seasonality and Persistence of Low-Frequency Atmospheric Circulation Patterns},
  journal = {Monthly Weather Review},
  year = {1987},
  volume = {115},
  number = {6},
  pages = {1083--1126},
  doi = {10.1175/1520-0493(1987)115<1083:CSAPOL>2.0.CO;2}
}

@article{Bi2023,
  author = {Bi, Kaifeng and Xie, Lingxi and Zhang, Hengheng and Chen, Xin and Gu, Xiaotao and Tian, Qi},
  title = {Accurate medium-range global weather forecasting with 3D neural networks},
  journal = {Nature},
  year = {2023},
  volume = {619},
  pages = {533--538},
  doi = {10.1038/s41586-023-06185-3}
}

@article{Dorigo2017,
  author = {Dorigo, Wouter A. and Wagner, Wolfgang and Albergel, Clement and Albrecht, Franz and Balsamo, Gianpaolo and Brocca, Luca and Chung, Dongryeol and Ertl, Martin and Forkel, Matthias and Gruber, Alexander and Haas, Elisabeth and Hamer, Paul D. and Hirschi, Martin and Ikonen, Jouni and de Jeu, Richard and Kidd, Richard and Lahoz, William and Liu, Yongwook and Miralles, Diego and Mistelbauer, Thomas and Nicolai-Shaw, Nicole and Parinussa, Robert and Pratola, Chiara and Reimer, Christoph and van der Schalie, Robin and Seneviratne, Sonia I. and Smolander, Thomas and Lecomte, Patricia},
  title = {{ESA CCI Soil Moisture for improved Earth system understanding: State-of-the art and future directions}},
  journal = {Remote Sensing of Environment},
  year = {2017},
  volume = {203},
  pages = {185--215},
  doi = {10.1016/j.rse.2017.07.001}
}

@article{Gneiting2005,
  author = {Gneiting, Tilmann and Raftery, Adrian E. and Westveld, Anton H. and Goldman, Tom},
  title = {Calibrated Probabilistic Forecasting Using Ensemble Model Output Statistics and Minimum CRPS Estimation},
  journal = {Monthly Weather Review},
  year = {2005},
  volume = {133},
  number = {5},
  pages = {1098--1118},
  doi = {10.1175/MWR2904.1}
}

@article{GneitingRaftery2007,
  author = {Gneiting, Tilmann and Raftery, Adrian E.},
  title = {Strictly Proper Scoring Rules, Prediction, and Estimation},
  journal = {Journal of the American Statistical Association},
  year = {2007},
  volume = {102},
  number = {477},
  pages = {359--378},
  doi = {10.1198/016214506000001437}
}

@article{Gronquist2021,
  author = {Gr{\"o}nquist, Peter and Yao, Chengyuan and Ben-Nun, Tal and Dryden, Nikoli and Dueben, Peter and Li, Shigang and Hoefler, Torsten},
  title = {Deep learning for post-processing ensemble weather forecasts},
  journal = {Philosophical Transactions of the Royal Society A},
  year = {2021},
  volume = {379},
  number = {2194},
  pages = {20200092},
  doi = {10.1098/rsta.2020.0092}
}

@article{Hersbach2000,
  author = {Hersbach, Hans},
  title = {Decomposition of the Continuous Ranked Probability Score for Ensemble Prediction Systems},
  journal = {Weather and Forecasting},
  year = {2000},
  volume = {15},
  number = {5},
  pages = {559--570},
  doi = {10.1175/1520-0434(2000)015<0559:DOTCRP>2.0.CO;2}
}

@article{Hersbach2020,
  author = {Hersbach, Hans and Bell, Bill and Berrisford, Paul and Hirahara, Shoji and Horanyi, Andras and Munoz-Sabater, Joaquin and Nicolas, Julien and Peubey, Carole and Radu, Raluca and Schepers, Dinand and Simmons, Adrian and Soci, Cornel and Abdalla, Saleh and Abellan, Xavier and Balsamo, Gianpaolo and Bechtold, Peter and Biavati, Gionata and Bidlot, Jean and Bonavita, Massimo and De Chiara, Giovanna and Dahlgren, Per and Dee, Dick and Diamantakis, Michail and Dragani, Rossana and Flemming, Johannes and Forbes, Richard and Fuentes, Manuel and Geer, Alan and Haimberger, Leopold and Healy, Sean and Hogan, Robin J. and Holm, El{\'i}as and Janiskov{\'a}, Marta and Keeley, Sarah and Laloyaux, Patrick and Lopez, Philippe and Lupu, Cristina and Radnoti, Gabor and de Rosnay, Patricia and Rozum, Iryna and Vamborg, Freja and Villaume, Sebastien and Th{\'e}paut, Jean-Noel},
  title = {The ERA5 global reanalysis},
  journal = {Quarterly Journal of the Royal Meteorological Society},
  year = {2020},
  volume = {146},
  number = {730},
  pages = {1999--2049},
  doi = {10.1002/qj.3803}
}

@article{Lam2023,
  author = {Lam, Remi and Sanchez-Gonzalez, Alvaro and Willson, Matthew and Wirnsberger, Peter and Fortunato, Meire and Alet, Ferran and Ravuri, Suman and Ewalds, Timo and Eaton-Rosen, Zach and Hu, Weihua and Merose, Alexander and Hoyer, Stephan and Holland, George and Vinyals, Oriol and Stott, Jacklynn and Pritzel, Alexander and Mohamed, Shakir and Battaglia, Peter},
  title = {Learning skillful medium-range global weather forecasting},
  journal = {Science},
  year = {2023},
  volume = {382},
  number = {6677},
  pages = {1416--1421},
  doi = {10.1126/science.adi2336}
}

@article{Pegion2019SubX,
  author = {Pegion, Kathy and Kirtman, Ben P. and Becker, Emily and Collins, Dan C. and Lajoie, Emerson and Burgman, Robert and Bell, Ray and DelSole, Timothy and Min, Dughong and Zhu, Yuejian and Li, Wei and Sinsky, Eric and Guan, Hong and Gottschalck, Jon and Metzger, E. Joseph and Barton, Neil P. and Achuthavarier, Deepthi and Marshak, Jelena and Koster, Randal D. and Lin, Hai and Gagnon, Normand and Bell, Michael and Tippett, Michael K. and Robertson, Andrew W. and Sun, Shan and Benjamin, Stanley G. and Green, Benjamin W. and Bleck, Rainer and Kim, Hyemi},
  title = {The Subseasonal Experiment ({SubX}): A Multimodel Subseasonal Prediction Experiment},
  journal = {Bulletin of the American Meteorological Society},
  year = {2019},
  volume = {100},
  number = {10},
  pages = {2043--2060},
  doi = {10.1175/BAMS-D-18-0270.1}
}

@article{Lang2026,
  author = {Lang, Simon and Alexe, Mihai and Clare, Mariana C. A. and Roberts, Christopher and Adewoyin, Rilwan and Bouall{\`e}gue, Zied Ben and Chantry, Matthew and Dramsch, Jesper and Dueben, Peter D. and Hahner, Sara and Maciel, Pedro and Prieto-Nemesio, Ana and O'Brien, Cathal and Pinault, Florian and Polster, Jan and Raoult, Baudouin and Tietsche, Steffen and Leutbecher, Martin},
  title = {{AIFS-CRPS}: Ensemble forecasting using a model trained with a loss function based on the Continuous Ranked Probability Score},
  journal = {npj Artificial Intelligence},
  year = {2026},
  volume = {2},
  pages = {18},
  doi = {10.1038/s44387-026-00073-7}
}

@article{Sun2018FIM,
  author = {Sun, Shan and Bleck, Rainer and Benjamin, Stanley G. and Green, Benjamin W. and Grell, Georg A.},
  title = {Subseasonal Forecasting with an Icosahedral, Vertically Quasi-Lagrangian Coupled Model. Part I: Model Overview and Evaluation of Systematic Errors},
  journal = {Monthly Weather Review},
  year = {2018},
  volume = {146},
  number = {5},
  pages = {1601--1617},
  doi = {10.1175/MWR-D-18-0006.1}
}

@inproceedings{Lipman2023,
  author = {Lipman, Yaron and Chen, Ricky T. Q. and Ben-Hamu, Heli and Nickel, Maximilian and Le, Matt},
  title = {Flow Matching for Generative Modeling},
  booktitle = {International Conference on Learning Representations},
  year = {2023},
  url = {https://arxiv.org/abs/2210.02747}
}

@inproceedings{Liu2023,
  author = {Liu, Xingchao and Gong, Chengyue and Liu, Qiang},
  title = {Flow Straight and Fast: Learning to Generate and Transfer Data with Rectified Flow},
  booktitle = {International Conference on Learning Representations},
  year = {2023},
  url = {https://arxiv.org/abs/2209.03003}
}

@techreport{Lorenz1956,
  author = {Lorenz, Edward N.},
  title = {Empirical Orthogonal Functions and Statistical Weather Prediction},
  institution = {Department of Meteorology, Massachusetts Institute of Technology},
  type = {Statistical Forecasting Project Scientific Report},
  number = {1},
  address = {Cambridge, Massachusetts},
  pages = {49},
  year = {1956},
  note = {AFCRC-TN-57-256}
}

@article{MaddenJulian1971,
  author = {Madden, Roland A. and Julian, Paul R.},
  title = {Detection of a 40--50 Day Oscillation in the Zonal Wind in the Tropical Pacific},
  journal = {Journal of the Atmospheric Sciences},
  year = {1971},
  volume = {28},
  number = {5},
  pages = {702--708},
  doi = {10.1175/1520-0469(1971)028<0702:DOADOI>2.0.CO;2}
}

@article{MathesonWinkler1976,
  author = {Matheson, James E. and Winkler, Robert L.},
  title = {Scoring Rules for Continuous Probability Distributions},
  journal = {Management Science},
  year = {1976},
  volume = {22},
  number = {10},
  pages = {1087--1096},
  doi = {10.1287/mnsc.22.10.1087}
}

@misc{MetOffice2022Heatwave,
  author = {{Met Office}},
  title = {Record high temperatures verified},
  year = {2022},
  url = {https://www.metoffice.gov.uk/about-us/news-and-media/media-centre/weather-and-climate-news/2022/record-high-temperatures-verified},
  note = {Accessed 2026-07-01}
}

@misc{CopernicusMarineSSS,
  author = {{Copernicus Marine Service}},
  title = {Multi Observation Global Ocean Sea Surface Salinity and Sea Surface Density},
  year = {2026},
  publisher = {Copernicus Marine Service},
  doi = {10.48670/moi-00051},
  url = {https://data.marine.copernicus.eu/product/MULTIOBS_GLO_PHY_S_SURFACE_MYNRT_015_013/description},
  note = {Product MULTIOBS\_GLO\_PHY\_S\_SURFACE\_MYNRT\_015\_013; accessed 2026-07-13}
}

@misc{C3SSoilMoisture,
  author = {{Copernicus Climate Change Service (C3S)}},
  title = {Soil Moisture Gridded Data from 1978 to Present},
  year = {2018},
  publisher = {Copernicus Climate Change Service Climate Data Store},
  doi = {10.24381/cds.d7782f18},
  url = {https://cds.climate.copernicus.eu/datasets/satellite-soil-moisture},
  note = {Accessed 2026-07-13}
}

@article{RaspLerch2018,
  author = {Rasp, Stephan and Lerch, Sebastian},
  title = {Neural Networks for Postprocessing Ensemble Weather Forecasts},
  journal = {Monthly Weather Review},
  year = {2018},
  volume = {146},
  number = {11},
  pages = {3885--3900},
  doi = {10.1175/MWR-D-18-0187.1}
}

@article{Reynolds2007,
  author = {Reynolds, Richard W. and Smith, Thomas M. and Liu, Chunying and Chelton, Dudley B. and Casey, Kenneth S. and Schlax, Michael G.},
  title = {Daily High-Resolution-Blended Analyses for Sea Surface Temperature},
  journal = {Journal of Climate},
  year = {2007},
  volume = {20},
  number = {22},
  pages = {5473--5496},
  doi = {10.1175/2007JCLI1824.1}
}

@book{RobertsonVitart2018,
  editor = {Robertson, Andrew W. and Vitart, Frederic},
  title = {Sub-seasonal to Seasonal Prediction: The Gap Between Weather and Climate Forecasting},
  publisher = {Elsevier},
  year = {2018}
}

@inproceedings{Ronneberger2015,
  author = {Ronneberger, Olaf and Fischer, Philipp and Brox, Thomas},
  title = {U-Net: Convolutional Networks for Biomedical Image Segmentation},
  booktitle = {Medical Image Computing and Computer-Assisted Intervention},
  year = {2015},
  pages = {234--241},
  doi = {10.1007/978-3-319-24574-4_28}
}

@article{Vitart2017,
  author = {Vitart, Frederic and Ardilouze, Constantin and Bonet, A. and Brookshaw, A. and Chen, M. and Codorean, C. and Deque, M. and Ferranti, L. and Fucile, E. and Fuentes, M. and Hendon, H. and Hodgson, J. and Kang, H.-S. and Kumar, A. and Lin, H. and Liu, G. and Liu, X. and Malguzzi, P. and Mallas, I. and Manoussakis, M. and Mastrangelo, D. and MacLachlan, C. and McLean, P. and Minami, A. and Mladek, R. and Nakazawa, T. and Najm, S. and Nie, Y. and Rixen, M. and Robertson, A. W. and Ruti, P. and Sun, C. and Takaya, Y. and Tolstykh, M. and Venuti, F. and Waliser, D. and Woolnough, S. and Wu, T. and Won, D.-J. and Xiao, H. and Zaripov, R. and Zhang, L.},
  title = {The Subseasonal to Seasonal (S2S) Prediction Project Database},
  journal = {Bulletin of the American Meteorological Society},
  year = {2017},
  volume = {98},
  number = {1},
  pages = {163--173},
  doi = {10.1175/BAMS-D-16-0017.1}
}

@article{WallaceGutzler1981,
  author = {Wallace, John M. and Gutzler, David S.},
  title = {Teleconnections in the Geopotential Height Field during the Northern Hemisphere Winter},
  journal = {Monthly Weather Review},
  year = {1981},
  volume = {109},
  number = {4},
  pages = {784--812},
  doi = {10.1175/1520-0493(1981)109<0784:TITGHF>2.0.CO;2}
}

@article{WheelerHendon2004,
  author = {Wheeler, Matthew C. and Hendon, Harry H.},
  title = {An All-Season Real-Time Multivariate MJO Index: Development of an Index for Monitoring and Prediction},
  journal = {Monthly Weather Review},
  year = {2004},
  volume = {132},
  number = {8},
  pages = {1917--1932},
  doi = {10.1175/1520-0493(2004)132<1917:AARMMI>2.0.CO;2}
}

@misc{CW3E2023CaliforniaAR,
  author       = {{Center for Western Weather and Water Extremes (CW3E)}},
  title        = {{California Department of Water Resources--CW3E Research and Operations Partnership: Highlights of a Family of Nine Atmospheric Rivers}},
  year         = {2023},
  howpublished = {CW3E event story map},
  url          = {https://cw3e.ucsd.edu/dec2022-jan2023_arstorymap/},
  note         = {Scripps Institution of Oceanography; accessed 2026-07-13}
}

@article{Otto2023PakistanFloods,
  author  = {Otto, Friederike E. L. and Zachariah, Mariam and Saeed, Fahad and Siddiqi, Ayesha and Kamil, Shahzad and Mushtaq, Haris and Arulalan, T. and AchutaRao, Krishna and Chaithra, S. T. and Barnes, Clair and Philip, Sjoukje and Kew, Sarah and Vautard, Robert and Koren, Gerbrand and Pinto, Izidine and Wolski, Piotr and Vahlberg, Maja and Singh, Roop and Arrighi, Julie and van Aalst, Maarten and Thalheimer, Lisa and Raju, Emmanuel and Li, Sihan and Yang, Wenchang and Harrington, Luke J. and Clarke, Ben},
  title   = {Climate change increased extreme monsoon rainfall, flooding highly vulnerable communities in {Pakistan}},
  journal = {Environmental Research: Climate},
  volume  = {2},
  number  = {2},
  pages   = {025001},
  year    = {2023},
  doi     = {10.1088/2752-5295/acbfd5}
}

@article{Ferro2014,
  author  = {Ferro, Christopher A. T.},
  title   = {Fair scores for ensemble forecasts},
  journal = {Quarterly Journal of the Royal Meteorological Society},
  volume  = {140},
  number  = {683},
  pages   = {1917--1923},
  year    = {2014},
  doi     = {10.1002/qj.2270}
}

@article{Richardson2001,
  author  = {Richardson, David S.},
  title   = {Measures of skill and value of ensemble prediction systems, their interrelationship and the effect of ensemble size},
  journal = {Quarterly Journal of the Royal Meteorological Society},
  volume  = {127},
  number  = {577},
  pages   = {2473--2489},
  year    = {2001},
  doi     = {10.1002/qj.49712757715}
}

\end{document}